\documentclass[a4paper,fleqn,usenatbib]{mnras}

\usepackage[T1]{fontenc}
\usepackage{ae,aecompl}

\usepackage{graphicx}	
\usepackage{amsmath}	
\usepackage{amssymb}	
\usepackage{txfonts}

\def\mean#1{\left< #1 \right>}
\graphicspath{{figures/}}
\title[Disk Polarization]{Disk Polarization From Both Emission and Scattering of Magnetically Aligned Grains: The Case of NGC 1333 IRAS4A1}

\author[H. Yang et al.]{
Haifeng Yang$^{1}$\thanks{E-mail: hy4px@virginia.edu},
Zhi-Yun Li$^{1}$,
Leslie W. Looney$^{2}$,
Erin G. Cox$^{2}$,
John Tobin$^{3}$, 
\and
Ian W. Stephens$^{4}$,
Dominque M. Segura-Cox$^{2}$
and Robert J. Harris$^{2}$
\\
$^{1}$Astronomy Department, University of Virginia, Charlottesville, VA 22904, USA\\
$^{2}$Department of Astronomy, University of Illinois at Urbana-Champaign, Urbana, IL 61801, USA\\
$^{3}$Leiden Observatory, Leiden University, P.O. Box 9513, 2000-RA Leiden, The Netherlands\\
$^{4}$Institute for Astrophysical Research, Boston University, Boston, MA 02215, USA
}

\date{Accepted XXX. Received YYY; in original form ZZZ}

\pubyear{2016}

\begin{document}
\label{firstpage}
\pagerange{\pageref{firstpage}--\pageref{lastpage}}
\maketitle

\begin{abstract}

Dust polarization in millimeter (and centimeter) has been mapped in
disks around an increasing number of young stellar objects. It is
usually thought to come from emission by magnetically aligned 
(non-spherical) grains, but can also be produced by dust scattering. 
We present a semi-analytic theory of disk polarization that includes 
both the direction emission and scattering, with an emphasis on 
their relative importance and how they are affected by the disk 
inclination. For face-on disks, both emission and scattering tend 
to produce polarization in the radial direction, making them 
difficult to distinguish, although the scattering-induced polarization 
can switch to the azimuthal direction if the incident radiation is 
beamed strongly enough in the radial direction in the disk plane. 
Disk inclination affects the polarizations from emission and 
scattering differently, especially on the major axis where, in 
the edge-on limit, the former vanishes while the latter reaches 
a polarization fraction as large as $1/3$. The polarizations 
from the two competing mechanisms tend to cancel each other on the 
major axis, producing two low polarization ``holes'' (one on each 
side of the center) under certain conditions. We find tantalizing 
evidence for at least one such ``hole'' in NGC1333 IRAS4A1, 
whose polarization observed at 8~mm on the 100~AU scale is 
indicative of a pattern dominated by scattering close to the 
center and by direction emission in the outer region. If true, 
it would imply not only that a magnetic field exists on the disk 
scale, but that it is strong enough to align large, possibly 
mm-sized, grains.  

\end{abstract}




\section{Introduction}

It is generally expected that magnetic fields play a crucial role 
in the dynamics and evolution of young star disks, through
magneto-rotational instability \citep{BH91} and
magneto-centrifugal disk wind (\citealt{BP82}; see 
\citealt{turner2014} and \citealt{armitage2015} for recent reviews). 
This expectation, based mostly on theoretical studies, provides 
a strong motivation to search for the putative disk field
observationally. 
To date, the observational effort has been
concentrated on detecting and characterizing the polarized dust
continuum emission, which has long been interpreted as coming 
from magnetically aligned grains \citep{lazarian2007, andersson2015}, 
using the Submillimeter
Array \citep[SMA;][]{hughes2009,rao2014} and Combined Array for Research 
in Millimeter-wave Astronomy \citep[CARMA;][]{hughes2013,stephens2014, segura-cox2015}. 
More recently, \cite{cox2015} opened 
a new front for this line of research by detecting dust 
polarization at 8 mm and 1 cm on the 100-AU scale around the 
protostar NGC1333 IRAS4A1 using The Karl G.
Jansky Very Large Array (VLA), as part of the VLA Nascent Disk 
and Multiplicity (VANDAM) survey (\citealt{tobin2015}; see also 
\citealt{liu2016}). If the detected (sub)mm and cm polarization 
is indeed produced by magnetically aligned grains, it would 
provide the long sought-after evidence that young stellar disks 
are magnetized, which is a pre-requisite for MRI and 
magneto-centrifugal disk winds to operate. 

However, \cite{kataoka2015a} recently discovered an alternative 
mechanism for producing polarized millimeter emission in disks 
that relies on dust scattering of anisotropic incident radiation 
rather than the alignment of asymmetric grains. \citet[Paper I hereafter]{yang2016}
showed that, in the best observed case of HL Tau 
disk \citep{stephens2014}, 
the polarization pattern is broadly consistent with that produced 
by dust scattering in an inclined disk \citep[see also][]{kataoka2015b}, 
although grain alignment cannot be ruled out completely, 
especially 
if the magnetic field structure of the disk is more complex than 
a purely toroidal configuration \citep{stephens2014}. If the 
dust scattering interpretation is correct, the grains responsible 
for the scattering in the HL Tau disk must be orders of magnitude
larger than the classical ISM size of 0.1~$\mu$m (at least several 
tens of microns; Paper I and \citealt{kataoka2015b}). The 
inferred (relatively large) grain size would add to other lines 
of evidence for substantial grain growth in protoplanetary disks 
(e.g., \citealt{perez2012,guidi2016}; see \citealt{testi2014}
for a recent review), which provides a first 
step toward planets. 

Whether large (non-spherical) grains can be aligned with respect to 
the magnetic field inside a protoplanetary disk remains uncertain. 
In the context of the currently favored mechanism for grain alignment 
through radiative torque, their magnetic moments may not be large 
enough to provide the fast precession needed (\citealt{lazarian2007}; although 
it depends on the disk field strength, which is uncertain), and their 
slow internal relaxation makes the alignment less efficient \citep{HL09}. 
More work is needed to address this important 
issue. In this paper, we will adopt the conventional view that the 
grains are aligned with respect to the magnetic field 
\citep{andersson2015}, at least to some extent in the disk, and treat the 
(currently uncertain) degree of alignment as a free
parameter\footnote{The parametrization is also needed because 
  of the uncertainty in the grain shape, which greatly affects 
  the degree of polarization but cannot be determined
  from the grain alignment theory.}. 
This treatment allows us to focus on the following question: how would 
the polarization pattern produced by direct emission from magnetically 
aligned grains be modified by scattering by the same aligned grains? It is a 
step beyond Paper I and \cite{kataoka2015a, kataoka2015b}, 
because the grains are no longer assumed to be spherical 
and the polarization from direct dust emission is included together
with that from scattering. Our goal 
is to delineate the conditions under which one of the two competing 
mechanisms would dominate over 
the other and vice versa, and to determine the composite polarization pattern
when both are important. This delineation of the parameter space and
the determination of polarization pattern will benefit the physical   
interpretation of disk polarization observations, especially those  
to be conducted with the Atacama Large Millimeter/submillimeter Array 
(ALMA). 

As a first step toward a comprehensive theory of disk polarization
including both emission and scattering from magnetically aligned 
grains, we will adopt the well-known ``electrostatic approximation'' 
to simplify the computation of the optical properties of non-spherical
grains. This approximation is discussed in \S~\ref{sec:electrostatic}, 
together with an {\it analytic} model to illustrate the relative 
importance of the scattering and direct emission in producing 
polarization, which turns out to depend 
strongly on the disk inclination. In \S~\ref{sec:disk}, we compute 
numerically the polarization patterns of a model disk produced by 
the scattering and direct emission individually and in combination, 
to illustrate the diverse outcomes of the competition 
between the two mechanisms, especially for disks of different 
inclinations. Our results are used to explain the polarization 
detected in NGC1333 IRAS 4A1 in \S~\ref{sec:application}. We 
discuss the implications of our results and their limitations in 
\S~\ref{sec:discussion}, and conclude in \S~\ref{sec:conclusion}.

\section{Competition between scattering and direct emission of
  non-spherical grains: analytic results}
\label{sec:electrostatic}

In order to determine how a non-spherical dust grain scatters light,
one needs to know how it interacts with an external electromagnetic
wave. The interaction can be very complicated in general, since 
each grain can be considered as a collection of polarizable parts, 
and each part responds to its local electric field inside the grain 
and may have a different polarization and phase. The grain-light 
interaction can in principle be 
treated numerically using, for example, the Discrete Dipole 
Approximation \citep[e.g.,][]{DF94}. However, such numerical 
treatments  
tend to be computationally expensive, and are not optimal for an 
initial exploration of the problem at hand: competition between 
the scattering and direct emission of non-spherical, magnetically-aligned 
grains in determining the polarization pattern. For such a purpose, we 
have decided to employ the well-known ``electrostatic approximation'' 
(e.g., \citealt{BH83}), which greatly 
simplifies the computation 
of the scattering cross sections without sacrificing the essential 
physics. The limitations of this approximation and its future 
refinements are discussed in section \S~\ref{sec:discussion} (see Fig.~\ref{fig:opacity}). Our
discussion below follows closely that in Chapter 5 of Bohren \& Huffman 1983, to which
we refer the readers for details.  

\subsection{Electrostatic approximation}

The basic requirement for the electrostatic approximation is that the
grain size is smaller than the wavelength of the external electromagnetic
wave. In such a case, the electric field varies little across the
grain, and the field can be approximated as having the same time dependence
throughout the region of interest.
The approximation simplifies the calculation of the polarization
of the (small) grain using the electrostatic equations with only
spatial derivatives. 

The scattering cross sections depend on both the size and shape of the 
dust grain. The grain shape is not well constrained. For illustration
purposes, we model the grain as an ellipsoid, for which analytic
solutions are available. For an ellipsoid composed of isotropic 
material with a complex dielectric constant $\epsilon$, the governing 
electrostatic equations can be solved analytically using ellipsoidal 
coordinates. The dielectrics will respond to the external field 
linearly and develop a dipole moment. Since the grains are not 
spherical, the polarizability $\bar{\alpha}$ that relates the electric dipole 
moment $\mathbf{p}$ induced in the grain to the external electric field 
$\mathbf{E}$  is not a single number but rather a matrix, i.e., 
$\mathbf{p} = \bar{\alpha} \mathbf{E}$. 
In a coordinate system with axes along the three principle axes of 
the dust grain, the polarizability matrix 
$\bar{\alpha}$ is diagonal, i.e.,  $\bar{\alpha} = \mathrm{diag}
\{\alpha_1, \alpha_2, \alpha_3\}$. Its diagonal element can be
expressed as: 
\begin{equation}
  \alpha_i = 4\pi r_e^3 \frac{\epsilon-1}{3+3L_i(\epsilon-1)},
  \label{eq:alpha_i}
\end{equation}
where $r_e$ is the radius of the sphere that has the same volume as the ellipsoid,  
and $L_i$ ($\rm i=1,2,3$) is a geometric parameter determined solely by the shape of
the grain, subjected to the constraint $L_1+L_2+L_3=1$. In the
simplest case of a spherical grain, $L_i$ is $1/3$. For an 
ellipsoidal grain, $L_i$ can be expressed as an integral that 
includes the length of the corresponding principle axis as a 
parameter. 
For an spheroid, which is an ellipsoid obtained by rotating an ellipse along
one of its principle axis, the integral can be done analytically. Following the
convention $L_1\le L_2\le L_3$ (which corresponds to the convention
for the semi-diameters $a_1 \ge a_2 \ge a_3$ and diagonal matrix 
elements $|\alpha_1| \ge |\alpha_2| \ge |\alpha_3|$), we have for a prolate 
spheroid ($a_1>a_2=a_3$):
\begin{equation}
  L_1 = \frac{1-e^2}{e^2}
  \left(-1+\frac{1}{2e}\mathrm{ln}\frac{1+e}{1-e}\right), \ 
  \qquad e^2 = 1-s^2,
  \label{eq:L_prolate}
\end{equation}
where $s=a_2/a_1 < 1$ is the axis ratio. The other two geometric
parameters are both equal to $(1-L_1)/2$. 

For an oblate spheroid ($a_1=a_2>a_3$), we have:
\begin{equation}
  \begin{split}
  &L_1 = \frac{g(e)}{2e^2}\left[\frac{\pi}{2}-\mathrm{tan}^{-1}g(e)\right] - \frac{g^2(e)}{2},\\
  &g(e)=\left(\frac{1-e^2}{e^2}\right)^{1/2},\qquad e^2=1-\frac{1}{s^2},
  \end{split}
  \label{eq:L_prolate}
\end{equation}
where the axis ratio is defined as $s=a_1/a_3>1$. The other two
geometric parameters are given by $L_2=L_1$ and $L_3=1-2L_1$.

As the external electric field varies over time, the dipole induced 
inside the grain also oscillates, which results in dipole radiation. 
It is straightforward to compute the scattering matrix and phase
matrix of the induced dipole radiation and, through the optical 
theorem, obtain the absorption cross section. The resultant 
scattering and absorption cross sections will be used to compute 
numerically the polarization due to direct emission and scattering 
of (small) ellipsoidal grains in a young star disk in
\S~\ref{sec:disk}. Before doing so, we will first illustrate the 
main features of the polarization produced by the scattering of 
non-spherical grains analytically in a limiting case, which will 
also allow us to compare with previous work and build 
physical intuition of how the scattering of non-spherical grains 
depends on the disk inclination, a focus of this investigation. 

\subsection{Inclination-induced polarization from scattering by 
   oblate grains}
\label{subsec:oblate1}

In Paper I, we showed that the disk inclination with
respect to the line of sight plays an important role in the
polarization produced by the scattering of spherical grains. 
The inclination-induced
polarization was illustrated analytically in the limiting case 
where the disk is geometrically thin and the incoming radiation 
to be scattered by the grains is locally isotropic in the disk 
plane (see their \S~2.2). Under these conditions, the 
polarization fraction of the scattered light by small spherical grains 
goes from zero for the face-on view to $1/3$ for the edge-on 
case. Here, we extend this analysis to oblate grains with the
semi-diameters $a_1 = a_2 > a_3$ \citep{HD95}; 
the case of prolate grains will be discussed in the Appendix~\ref{sec:imp&geo}.  

To be specific, let us consider the polarization of the light
scattered by oblate grains at a location $O$ inside a disk 
that is inclined with respect to the line of sight by an 
angle $i$ ($i=0^\circ$ corresponds to the face-on 
case). We will adopt a Cartesian coordinate system centered on 
the location $O$, with the $x$-axis pointing radially away from 
the center of the disk, and $y$-axis tangential to the circle 
in the disk plane that is centered at the origin and passes through 
the point $O$. For 
simplicity, we assume that the disk magnetic field is purely 
toroidal, so that the only non-zero component is along the 
$y$-direction. In the case of perfect grain alignment, the y-axis is 
also the direction of the minor axis of the oblate grain 
(with the smallest semi-diameter $a_3$). The $z$-axis of the 
coordinate system is perpendicular to the disk plane. 
In this coordinate system, the polarizability is diagonal: 
$\bar{\alpha} = \mathrm{diag}\left\{\alpha_x, \alpha_y, 
\alpha_z\right\}$, with $\alpha_x = \alpha_z \equiv \alpha_1$, 
$\alpha_y \equiv \alpha_3$ and $|\alpha_1|>|\alpha_3|$. 
We let the $x$-axis lie in the plane of the sky, so that the line of 
sight to the location $O$ of interest is perpendicular to the 
$x$-axis and is thus in the $yOz$ plane. In this coordinate 
system, the disk inclination angle $i$ is simply the angle between 
the $z$ axis and the line of sight, and the $x$-axis is along the 
major axis of the inclined disk projected in the plane of the sky. 
For this initial analysis, we focus on the disk locations on the major 
rather than the minor axis for two reasons. First, the polarization 
produced by direct emission from the oblate 
grains on the minor axis is independent of the inclination 
angle because these grains are aligned with the (toroidal) 
magnetic field in such a way that they always appear  
``edge-on'' to the observer. More importantly, the polarization 
pattern is expected to be simpler on the minor axis because both 
direct emission and scattering there tend to produce polarization 
along the minor axis (although not always, see Fig.~\ref{fig:ill}), so 
that they generally add to, rather than cancel, each other. 

Our goal is to determine the polarization properties of the light that
is scattered into our line of sight. In general, the Stokes 
parameters of the scattered light ($I_s$, $Q_s$, $U_s$ and $V_s$) 
are related to those of the incident radiation ($I_i$, $Q_i$, $U_i$
and $V_i$) through a 16-element scattering matrix (see \citealt{BH83},
p65). We assume that the 
incident light is non-polarized (i.e., $Q_i=U_i=V_i=0$), so 
that only 4 of the matrix elements are relevant, namely: $I_s\propto 
S_{11} I_i$, $Q_s\propto S_{21} I_i$, $U_s\propto S_{31} I_i$, and 
$V_s \propto S_{41} I_i$.  We assume further that the  
incident radiation to be scattered at the location $O$ is 
confined in the disk plane (i.e., the thin (dust) disk approximation), 
so that its direction is uniquely described by 
the azimuthal angle $\phi$ from the $x$-axis. In the limiting case
that the incident radiation is independent of the azimuthal angle 
$\phi$, it is straightforward to average the scattering 
matrix elements over $\phi$, which yields the following 
results:
\begin{equation}
  \mean{S_{11}} =
  \frac{1}{2}\left(\frac{k^3}{4\pi}\right)^2\left(|\alpha_1|^2 \sin^2 i
  +\frac{1}{2}|\alpha_3|^2 \cos^2 i +\frac{1}{2}|\alpha_1|^2\right),
  \label{eq:S11}
\end{equation}
\begin{equation}
  \mean{S_{21}} =
  -\frac{1}{2}\left(\frac{k^3}{4\pi}\right)^2\left(|\alpha_1|^2
  \sin^2 i
  +\frac{1}{2}|\alpha_3|^2 \cos^2 i -\frac{1}{2}|\alpha_1|^2\right),
  \label{eq:S21}
\end{equation}
where $k=2\pi/\lambda$ is the wave-number of the scattered light. 
In addition, $\mean{S_{31}}=\mean{S_{41}}=0$, as expected from the
symmetry of the problem. It means that the scattered light will
be polarized either in the $x$-direction or perpendicular to it,
and that there is no circular polarization. Since $\mean{S_{11}}$ and 
$\mean{S_{21}}$ are essentially the differential scattering cross
sections for the Stokes parameter $I$ and $Q$, respectively, the
degree of polarization of the scattered light is simply given by
their ratio:
\begin{equation}
  p_{\rm sca} = \frac{\mean{S_{21}}}{\mean{S_{11}}} = 
  \frac{|\alpha_1|^2-2|\alpha_1|^2 \sin^2 i -|\alpha_3|^2 \cos^2 i}
  {|\alpha_1|^2+2|\alpha_1|^2 \sin^2 i +|\alpha_3|^2 \cos^2 i},
  \label{eq:p_sca}
\end{equation}
which can be either positive or negative; a positive (negative)
$p_{sca}$ means that the polarization direction is parallel 
(perpendicular) to the $x$-axis in the plane of the sky. 

In order to obtain numerical values for $p_{sca}$, a grain model is
needed to calculate the values of $\alpha_1$ and $\alpha_3$. This 
will be done in the next subsection. Here, we will make a couple 
of interesting points that are independent of the detailed grain 
properties. First, since $|\alpha_1|>|\alpha_3|$ for oblate 
grains, we have $p_{sca}>0$  in the face-on case with $i=0^\circ$,
which means that the scattered light will be polarized in the 
$x$-direction. This is different from the case of spherical 
grains, where the polarization in the face-on case is zero. The
difference makes physical sense because, for non-spherical grains, 
the scattering cross sections for incident light coming from 
different directions are no longer the same. In particular, for 
oblate grains with the short axis aligned with the $y$-axis, light
propagating along the $y$-direction will be scattered more efficiently 
into our line of sight, producing polarization in the $x$-direction. 
The degree of polarization will depend on the degree of the grain 
non-sphericity, as we show below. Second, in the opposite limit of 
edge-on view ($ i= 90^\circ$), we have $p_{sca} = - 1/3$. This is 
expected because, when viewed edge-on, the grain is axisymmetric 
with respect to the line of sight. It means that the polarization 
in this limit is determined completely by the inclination effect, 
which is known to produce a fractional polarization of $1/3$ 
perpendicular to the $x$-axis in the plane of the sky (i.e., 
along the minor axis of the inclined disk, Paper I). In 
the limit $\alpha_1=\alpha_3$, we have $p_{sca} = 
-\sin^2 i /(2+\sin^2 i )$, which recovers the previous
analytic results for spherical grains\footnote{Note that the Stokes
  parameters in Paper I were defined in a plane-of-sky
  coordinate system $x'$-$y'$, with $x'$ along the minor axis of the
  inclined disk. In this paper, the $x$-axis lies in the plane of 
  the sky and is along the major axis of the disk. This difference 
introduces a sign difference between these two results}. 

\subsection{Competition between scattering and direct emission}
\label{subsec:oblate2}

In this subsection, we will compute the polarization from the
scattering of oblate grains at a location $O$ in an inclined disk 
adopting a specific grain model. The model allows us to determine 
diagonal elements of the polarizability matrix, $\alpha_1$ and 
$\alpha_3$, and, through Equation~\ref{eq:p_sca}, the degree of 
polarization, $p_{sca}$. The polarization from scattering will 
be compared with that from the direct emission from the same 
magnetically aligned oblate grains at the location $O$ (with 
the shortest axis along the $y$-direction). To determine the 
latter, the absorption cross sections along the major axis 
of the inclined disk in the plane of the sky, the $x$-axis, 
and the minor axis (denoted by $y'$ hereafter), are needed. They 
are related to the polarizability, especially the imaginary part, through the 
optical theorem:  
\begin{equation}
      \sigma_{\rm abs, x} = k\ \mathrm{Im}\left[\alpha_1\right],
\label{eq:sigma_abs_x}
\end{equation}
\begin{equation}
      \sigma_{\rm abs, y'} = k\ \mathrm{Im}\left[\alpha_3\cos^2 i
        +\alpha_1\sin^2 i \right],
  \label{eq:sigma_abs_y'}
\end{equation}
where $\mathrm{Im}[x]$ stands for the imaginary part of any variable $x$. 
These absorption cross sections 
yield the following degree of polarization for the direct emission:
\begin{equation}
  p_{\rm abs} = \frac{\mathrm{Im}\left[\alpha_1-\alpha_3\right]\cos^2
    i }
  {\mathrm{Im}\left[\alpha_3\cos^2 i +\alpha_1(1+\sin^2 i )\right]}. 
  \label{eq:p_abs}
\end{equation}

\begin{figure}
  \includegraphics[width=0.49\textwidth]{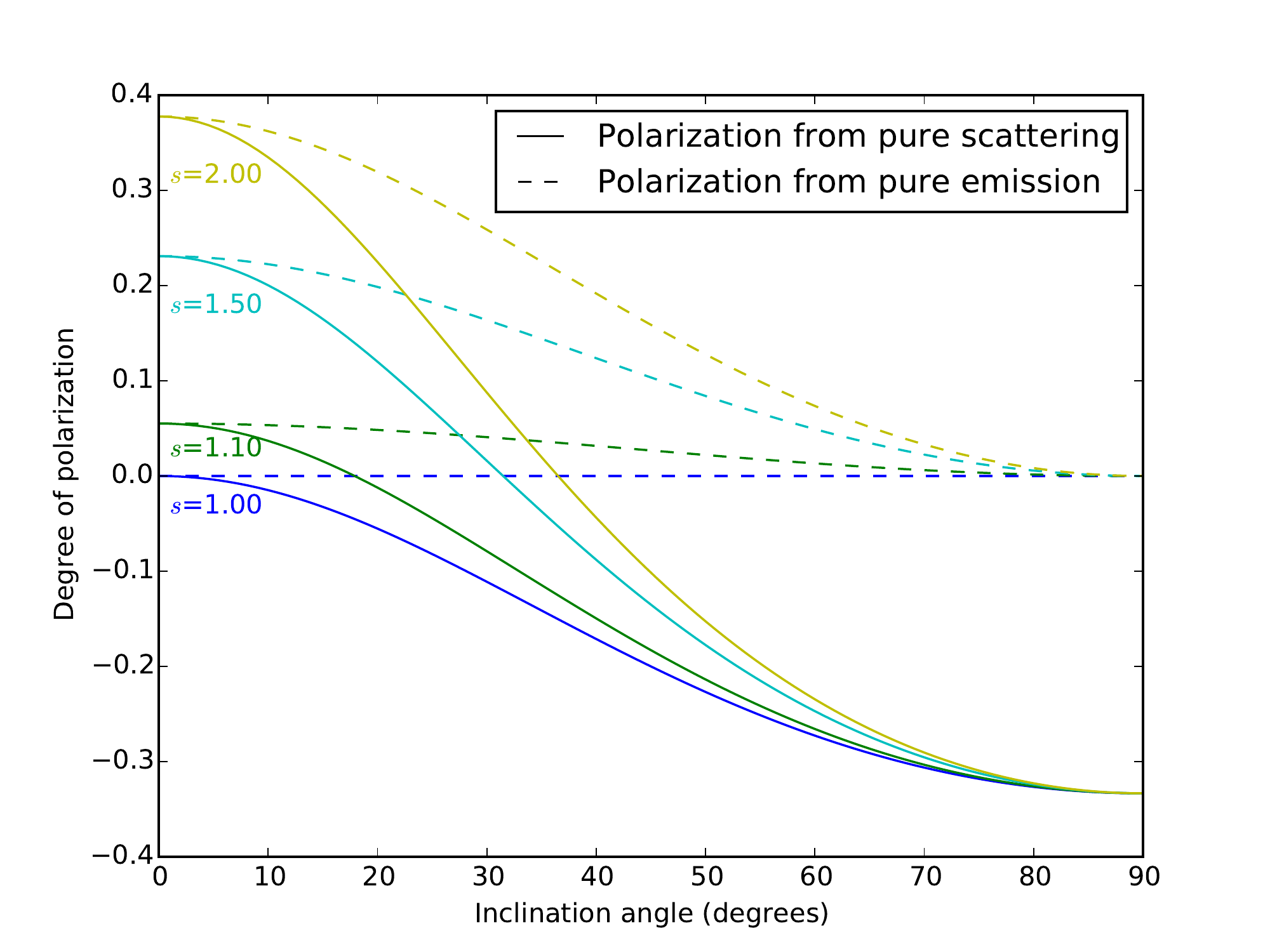}
  \caption{Degree of polarization at a location on the disk major axis 
   for scattered light ($p_{sca}$,
    solid lines) and direct emission ($p_{abs}$, dashed) for oblate
    grains with representative axis ratio $s=1.0$, $1.1$, $1.5$, and
    $2.0$ as a function of disk inclination angle $i$, assuming 
    perfect grain alignment. Note that $p_{sca}$ and $p_{abs}$ start
    from the same positive value at $i=0^\circ$ (the face-on limit), 
    but decrease to $-1/3$ and $0$, respectively, as the edge-on
    ($i=90^\circ$) limit is approached. 
}
  \label{fig:p_pol}
\end{figure}

We follow \cite{kataoka2015a} in adopting the grain model of
\cite{pollack1994}, where grains are composed of silicate (8\% in 
volume), water ice (62\%) and organics (30\%). This type of dust 
grains has a complex dielectric constant of $\epsilon = 3.78+0.04j$ (where
$j$ is the imaginary unit $\sqrt{-1}$) at $1$ mm. In Fig. \ref{fig:p_pol}, we plot the degree 
of polarization for scattered light
and direction emission, $p_{sca}$ and
$p_{abs}$, for several representative values of the axis ratio of the
oblate grain, $s=1.0$, $1.1$, $1.5$ and $2.0$, as a function of the 
disk inclination angle $i$. Several features are immediately
apparent. First, in the limit of spherical grains with $s=1.0$,  we
recover the well known (analytic) results that the direct emission is
not polarized, and the polarization from scattering is along the minor 
axis, with a polarization fraction that goes from zero to $1/3$ as 
the inclination angle $i$ increases from $0^\circ$ to $90^\circ$. Second, 
as anticipated analytically in the last subsection, the polarization 
of the light scattered by the oblate grains aligned with a toroidal 
magnetic field (along the $y$-direction) is along the $x$-axis (with 
a positive $p_{sca}$) in the face-on case. As the inclination angle
increases, the polarization along the major (or $x$-) axis is
gradually weakened by that from the polarization induced by the
inclination, which is along the minor (or $y'$-) axis. At a critical
inclination angle $i_c$, the polarization direction switches 
from the major axis to the minor axis; the angle $i_c$ increases
with the axis ratio $s$. In all cases, the scattering degree of
polarization asymptotes to the limiting value $p_{sca}=-1/3$ as the 
inclination angle $i$ approaches $90^\circ$, as we showed
analytically above. Third, the polarization of the direct emission 
by the aligned oblate grains is always along the major (or $x$-) 
axis of the disk, which is the direction of the long axis of 
the grain. The polarization degree $p_{abs}$ peaks in the face-on 
case, where the grain appears most elongated to the observer. 
Interestingly, the peak value is exactly the same as that of 
the scattering polarization degree $p_{sca}$ in the face-on 
case, which can be proven analytically for oblate grains. 
Lastly, the emission polarization degree $p_{abs}$ decreases smoothly 
with the inclination angle $i$, reaching zero in the edge-on 
limit, when the oblate grains appear circular to the observer and 
thus there is no preferred direction for polarization. The vanishing 
of $p_{abs}$ as $i \rightarrow 90^\circ$ means that the polarization
will be dominated sooner or later by scattering, as long as the
inclination angle $i$ is large enough.   

%
%
 
The relative contribution of scattering and direct emission to the 
polarization depends on not only the degree of polarization ($p_{\rm sca}$
and $p_{\rm abs}$), but also the ratio of $\sigma_{\rm sca}J_\nu$ and 
$\sigma_{\rm abs}B_\nu$, where $J_\nu$ is mean intensity at the 
location under consideration, $B_\nu$ is the local source 
function for thermal dust emission, and $\sigma_{\rm sca}$ and
$\sigma_{\rm abs}$ are the scattering and absorption cross sections. 
The ratio $J_\nu/B_\nu$ depends 
on the detailed disk model and temperature structure, while the 
ratio of scattering and absorption cross sections, $\sigma_{\rm
  sca}/\sigma_{\rm abs}$, depends on the dust composition and
especially grain size. Roughly speaking, the cross section ratio 
is of the order $(2\pi r_e/\lambda)^3$. In order for the scattering 
to be competitive, the grain size $r_e$ cannot be much smaller than
the wavelength $\lambda$. 
On the other hand, the electrostatic approximation that we
adopted is valid only when the grain is relatively small compared 
to the wavelength. As we show in \S~\ref{sec:discussion} below, the
scattering opacity exceeds the absorption opacity as long as the
grains are bigger than $\sim 0.05\lambda$, while the electrostatic
approximation remains valid for grain sizes up to $\sim 0.2 
\lambda$. For larger grains, the scattering opacity remains larger
than the absorption opacity, but their optical properties need 
to be computed numerically; we postpone such a treatment to a 
future investigation.
In what follows, we will leave the ratio 
$\sigma_{\rm abs}B_\nu/\sigma_{\rm sca}J_\nu$ as a free parameter, and
explore the parameter space where the polarization from scattering 
becomes important relative to that from direction emission. 

\begin{figure}
  \includegraphics[width=0.49\textwidth]{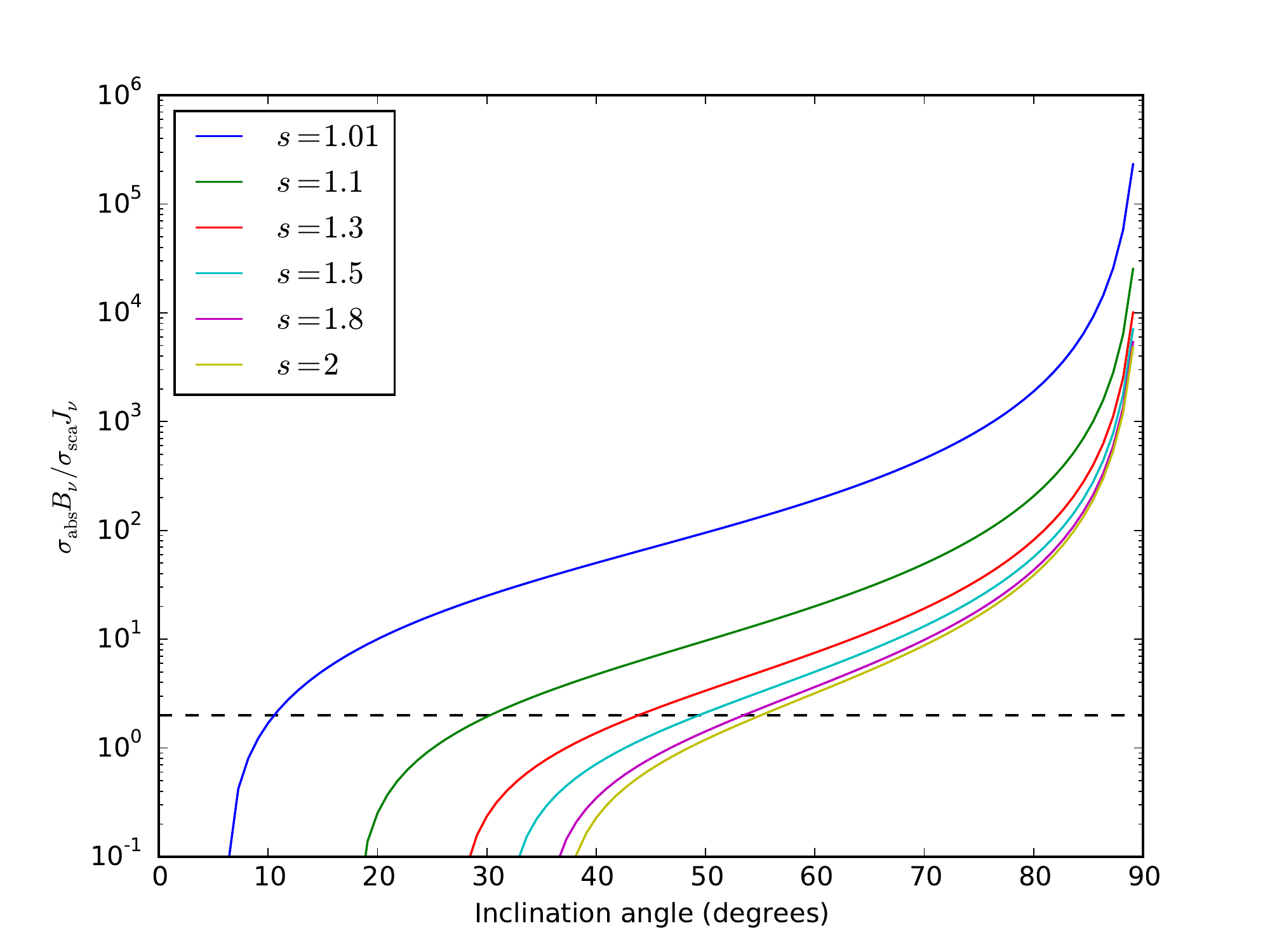}
  \caption{Transition lines that divide the parameter space where the
    polarization is 
    dominated by direct emission (to the upper left of each line) from
    that dominated by scattering (the lower right), for 6 
    representative values of the grain axis ratio $s=1.01$, 1.1,
    1.3, 1.5, 1.8 and 2.0. The horizontal line marks $\sigma_{\rm abs}
    B_\nu/\sigma_{\rm sca}J_\nu=2$, the fiducial value obtained in 
    the flared disk model of \protect\cite{CL07}.}
  \label{fig:rratio}
\end{figure}

Since the polarization from direct emission at a location on the major 
axis is always along the major axis (for a purely toroidal magnetic 
field), one way to measure the importance of the scattering is to 
determine the transition inclination angle $i_t$ beyond 
which the polarization is forced to align with the minor axis
instead. In Fig.~\ref{fig:rratio}, we plot the angle $i_t$ as a function 
of the ratio $\sigma_{\rm abs}B_\nu/\sigma_{\rm sca}J_\nu$ for a
representative set of values for the axis ratio $s$. Roughly 
speaking, for each value of $s$, the polarization is dominated by
direct emission in the parameter space to the upper left of the
corresponding curve, and by scattering to the lower right of the 
curve. Also shown in the plot is the fiducial value of $\sigma_{\rm abs}
B_\nu/\sigma_{\rm sca}J_\nu=2$ derived in the flared disk model of 
\cite{CL07}. For this fiducial
value, the polarization is dominated by scattering for $i$
greater than approximately $55^\circ$ as long as the grain 
axis ratio is not too extreme ($s < 2$, see \citealt{HD95}). 
For larger ratios of 
$\sigma_{\rm abs}B_\nu/\sigma_{\rm sca}J_\nu$, a more extreme
inclination is required for the scattering to become dominant, 
unless the grains are nearly spherical (i.e., with $s$ close 
to 1). In what follows, we will evaluate this ratio 
self-consistently with the help of a specific disk model.  
The effects of two potential complications, imperfect 
grain alignment and non-oblate grain shape, are discussed in the
Appendix~\ref{sec:imp&geo}. 
%
%

\section{Competition between scattering and direct emission in young
  star disks: numerical examples}
\label{sec:disk}

So far, we have limited our (analytic) discussion of the interplay
between the polarizations produced by non-spherical grains through 
scattering and direct emission to the limiting case where the 
incident radiation field is both planar and isotropic in 
the disk plane.  While the planar approximation is usually a good 
one, especially for large grains that tend to settle to the disk 
mid-plane, the isotropic assumption is adopted mainly for the 
purposes of illustrating the competition between scattering and 
emission as simply as possible. In this section, we will relax this
assumption with the help of a specific model 
for the disk structure, which enables a self-consistent computation 
of the angular distribution of the incident radiation field, as 
done in Paper I. More importantly, the disk model allows for a 
determination of the polarization pattern over the entire 
disk, which is needed for direct comparison with spatially resolved 
polarization observations, especially those with ALMA. 
We will keep the ``thin-disk'' approximation adopted in Paper I,  
which has been shown to greatly speed up the computation 
of the scattering-induced polarization in an inclined disk by 
spherical grains without compromising the essential physics of 
the problem. Our treatment here is essentially a generalization of 
Paper I to the case of non-spherical grains, where both 
scattering and direct emission contribute to the polarization. 
It turns out that the
combined polarization pattern resembles that observed recently in 
NGC1333 IRAS4A with VLA at 8 mm and 1 cm \citep{cox2015}. 
The application of our
results to this specific source will be discussed in 
\S~\ref{sec:application}.  

\subsection{Problem setup} 

%
%
We will compute the polarizations due to the direct thermal emission 
and scattering by non-spherical grains separately. The former can 
be done through straightforward integration along each line of 
sight once the grain properties, magnetic field configuration, and 
degree of grain alignment are specified. The latter is more 
complicated because, along each line of sight, it involves 
the computation of the incident radiation field to be scattered 
at all locations and the integration of the scattered light. To 
treat the scattering-induced polarization, we will adopt the 
same basic problem setup as in Paper I (see \S~2.1 there  
for details). Particularly important for their formulation of the 
scattering problem is the assumption that the disk is both 
geometrically and optically thin. This simplification enabled us to 
relate the source function of the 
radiation scattered into the line of sight at any target location  
${\bf r}$ on the (thin, inclined) disk to the column density and 
temperature at a source location ${\bf r}_1$ (which supplies the 
photons to be scattered at ${\bf r}$), $\Sigma({\bf r}_1)$ and 
$T({\bf r}_1)$, through their equations (6)-(7), which are 
reproduced here for easy reference: 
\begin{equation}
  S\approx \frac{2\nu^2 k \kappa_{\rm abs}}{c^2 \sigma_s}
  \int_0^{2\pi}d\phi \frac{d\sigma}{d\Omega}\Lambda(\mathbf{r}, \phi)
  \label{eq:Sinfty}
\end{equation}
where $\nu$ is the frequency of the scattered light, $k$ is the Boltzmann
constant, $\kappa_{\rm abs}$ the absorption opacity, $c$ the speed of
light, $\sigma_s$ the solid angle-integrated (total) scattering cross
section, $d\sigma/d\Omega$ the differential scattering cross section, 
and the quantity $\Lambda(\mathbf{r}, \phi)$ is an integral along a
straight line on the disk that passes through the target location 
${\bf r}$ along a constant azimuthal angle $\phi$:  
\begin{equation}
  \Lambda(\mathbf{r}, \phi) \equiv \int_{H}^{\infty} dl
  \frac{\Sigma({\bf r}_1)T({\bf r}_1)}{l}, 
  \label{eq:Lambda}
\end{equation}
where $H$ is the local disk scale-height at ${\bf r}_1$ and $l$ is the
separation between the target and source locations, ${\bf r}$ and 
${\bf r}_1$.  

%
%
In the simpler case of (small) spherical grains considered previously 
in Paper I, the differential scattering cross section 
$d\sigma/d\Omega$ in equation~(\ref{eq:Sinfty}) is simply given by 
Rayleigh scattering. For non-spherical grains, there are two
potential complications. The first is that the incident  
radiation to be scattered at a given location is already polarized 
before scattering because it is emitted by non-spherical grains. 
In principle, one needs to determine the polarization state of the
incident radiation carefully, taking into account of the grain
orientation at each source location ${\bf r}_1$ along the line 
of integration in equation~(\ref{eq:Lambda}). 
For simplicity, we shall assume that the incident light is 
unpolarized before scattering. This approximation should 
not change the polarization produced by scattering
qualitatively,
as explained in the Appendix \ref{sec:unpolarized}.

The second complication is that, for non-spherical grains, the
scattering matrix (see equations~\ref{eq:S11} and \ref{eq:S21}), 
which determines the differential cross section $d\sigma/d\Omega$ in 
equation~(\ref{eq:Sinfty}), 
will depend on two angles, the incident radiation direction and 
line of sight direction in the frame of the dust grains, rather
than a single scattering angle, as it is in Rayleigh scattering.
These matrix elements can be computed easily once the grain properties
and degree of grain alignment are specified. For illustrative 
purposes, we will adopt the same grain model of \cite{kataoka2015a}
used in the last section (\S~\ref{sec:electrostatic}) and 
assume that the grains are oblate spheroids perfectly aligned 
with a purely toroidal magnetic field in the disk; grains of other 
shapes (e.g., prolate) and/or imperfectly aligned should produce
qualitatively similar results after averaging around the field
direction (see Appendix~\ref{sec:imp&geo}). We adopt 
a volume-equivalent radius $r_e=100$~$\mu$m to maximize the effects 
of the scattering of radiation at 1~mm wavelength and a rather 
large axis ratio of $s=1.5$, so that the direct emission is 
significantly polarized. Other choices of $r_e$ and $s$ would not
change the polarization patterns produced by scattering and direct 
emission individually, but will affect their relative importance in
a simple way: increasing $r_e$ ($s$) tends to make scattering (direct
emission) more important.   

\subsection{Numerical examples of disk polarization pattern from both
  scattering and emission}
%
%
%
%
%
For our numerical examples, we adopt the column density distribution of
the viscous disk model of \cite{pringle1981}:
\begin{equation}
  \Sigma(R) = \Sigma_0 \left(\frac{R}{R_c}\right)^{-\gamma}
  \mathrm{exp}\left[{-\left(\frac{R}{R_c}\right)^{2-\gamma}}\right],
  \label{eq:vis_disk}
\end{equation}
which is often used for modeling disk continuum observations 
\citep[e.g.,][]{testi2014,kwon2015}. 
The prescribed disk profile has an inner part with a power-law
distribution and an outer part dominated by an exponential cutoff. 
Most observed disks have an inferred value of the power index 
$\gamma$ between $\sim 0.5$ and $\sim 1$ \citep{andrews2009, segura-cox2016}.  
We have experimented with different values of $\gamma$ in this 
range and found similar polarization patterns. Only the results 
for the $\gamma=0.5$ case will be shown below.

The size of the model disk is set by the characteristic radius
$R_c$. It provides an overall scaling for the polarization 
pattern, but does not change the pattern itself. For definitiveness,
we choose $R_c=50$~AU, and truncate the disk beyond an outer radius 
$R_{\rm out}=3 R_c=150$~AU. The inner radius of the disk is set 
to $R_{\rm in}=1$~AU in order to prevent the column density from going to
infinity at the origin. For the temperature profile, we adopt the
simple prescription
\begin{equation}
T(R)=T_0 \left(\frac{R}{R_c}\right)^{-1/2},
\label{eq:T}
\end{equation}
which is approximately valid for disks heated by the central stellar
radiation \citep[e.g.,][]{hartmann1998}.
We will assume the radiation is in the Rayleigh-Jeans regime and all the
intensities will be presented in unit of the Planck function $B_\nu(T_0)$; the 
dimensionless intensities are independent of $T_0$. As a concrete 
illustrative example, we set the scale
factor for the total (gas and dust) column density to 
$\Sigma_0=17\rm\,g/cm^2$ (with a gas-to-dust-ratio of 
100), so as to prevent the optical depth for direct emission 
from becoming too large, especially at small radii,  on the one 
hand and to make the optical depth for scattering large enough 
that the scattering can compete with direct emission in producing 
polarization on the other. The key parameter that we will focus on is 
the inclination angle $i$, which is expected to change the 
balance between the polarization produced by scattering and 
that by direct emission, based on the analytic results described 
in \S~\ref{sec:electrostatic}. 

\begin{figure*}
  \includegraphics[width=0.92\textwidth]{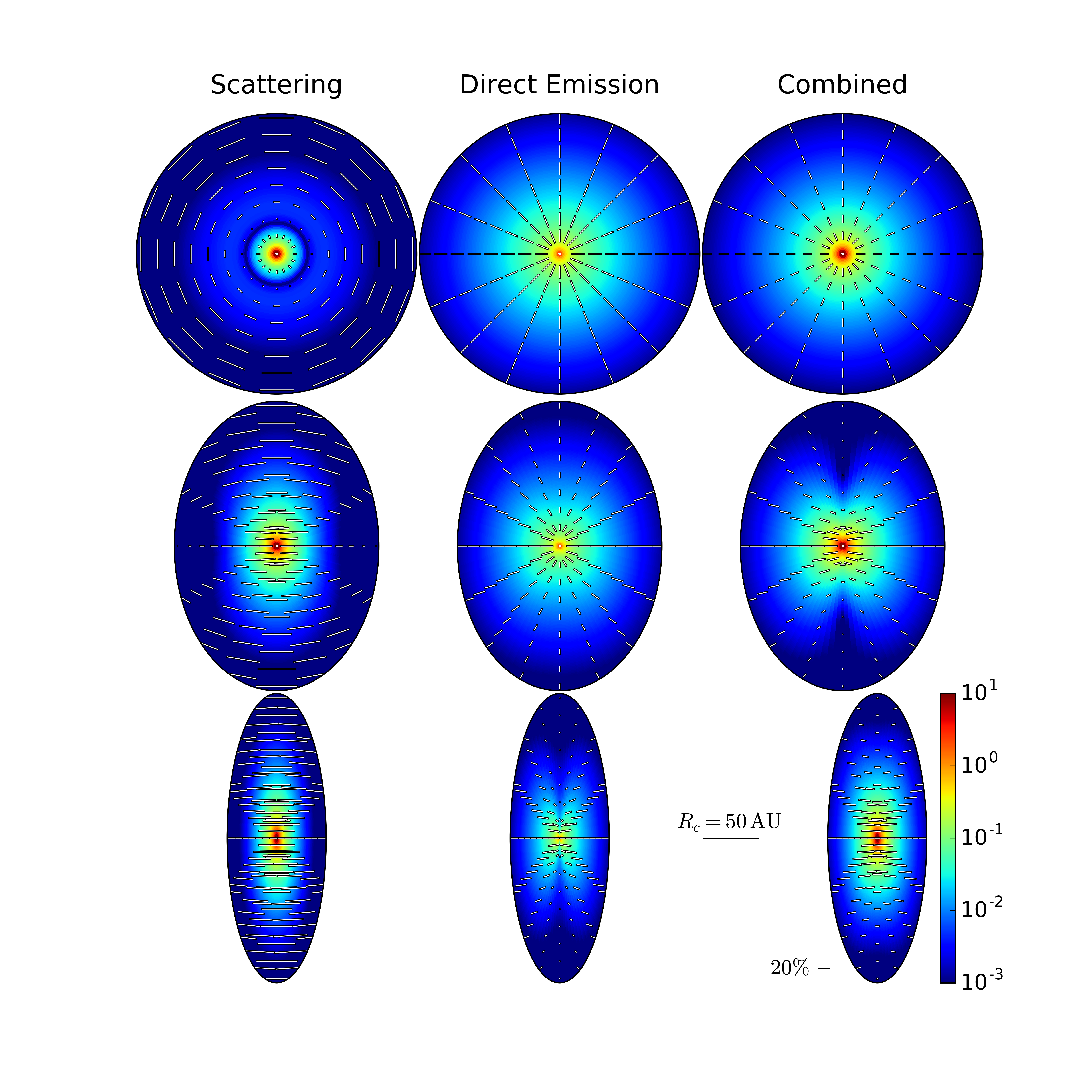}
  \caption{Polarized intensity (in units of $B_\nu(T_0)$, color map)
    and polarization vectors with length proportional to polarization
    fraction for scattering only (left panels), emission only (middle),
    and the two combined (right), for three inclinations $i=0^\circ$
    (upper panels), $45^\circ$ (middle), and $70^\circ$ (lower). 
}
  \label{fig:ill}
\end{figure*}

We will start with the simplest, face-on case ($i=0^\circ$), which is
free of any disk inclination effect. In this case, the polarization
pattern for the direct emission from the oblate grains that are perfectly
aligned with a purely toroidal magnetic field is trivial: the
polarization direction is radial everywhere (see upper-middle panel of 
Fig.~\ref{fig:ill}). The pattern for the
scattered light is more structured. The polarization direction is
radial inside a radius of $\sim 20$~AU (this radius depends on 
the disk mass and temperature distributions), and becomes azimuthal 
outside (see upper-left panel). 
This is very different from the pattern in the case of
spherical grains (see the top-left panel of Fig.~2 of Paper I), 
where the polarization direction is azimuthal everywhere,
including at small radii, where the polarization fraction is 
small, because the incident radiation field at these radii 
is more or less isotropic in the disk plane. In contrast, for
non-spherical grains, the scattered light can be significantly
polarized even for (planar) isotropic incident radiation, as we
demonstrated analytically in the last section (see 
Fig.~\ref{fig:p_pol}). Since the oblate grains are aligned 
with their shortest axes along the azimuthal (B-field) 
direction, incident light coming from the radial direction 
(with an electric field ${\bf E}$ along the azimuthal direction) 
is scattered less efficiently than that from the azimuthal 
direction (with ${\bf E}$ along the radial direction), leading 
to polarization along the radial direction at small radii 
where the incident radiation field in the disk plane is more 
or less isotropic. As the radius increases, the incident 
radiation field becomes more beamed in the radial direction,  
which leads to the polarization along the azimuthal direction 
in the outer part of the disk. Indeed, the incident radiation 
near the outer edge of the disk shown in Fig.~\ref{fig:ill} is 
so beamed in the radial direction that the polarization fraction 
iof the scattered light s more than 50\%. 

Despite the high polarization fraction, the polarized intensity of the 
scattered light is relatively low in the outer part in this 
particular example, so that the polarization of the combined light 
from both direct emission and scattering is in the radial 
direction everywhere (see the upper right panel). The radial 
polarization pattern does not mean that the direct emission 
dominates the polarization everywhere. Indeed, close to the center, 
the polarization is dominated by scattering\footnote{The exact size of
  the scattering dominated central region depends on the disk
  structure and dust properties, and will require more elaborate
  models to determine if the region becomes optically thick.}. 
This illustrates the 
potential danger of automatically identifying radial polarization 
with the direct emission from grains aligned with a toroidal 
magnetic field in a face-on disk. Other pieces of information, 
such as grain properties and disk radiation field, are needed to 
help determine unambiguously which polarization mechanism 
dominates. 
%
%

%
%

As the angle $i$ increases, the inclination-induced polarization 
in the scattered light becomes more important, which reduces the  
difference between the spherical and non-spherical grain cases 
(compare the lower-left panel of Fig.~2 of Paper I with the
middle-left panel of Fig.~\ref{fig:ill} for the $i=45^\circ$ 
case). In particular, in the inner part of the disk where the 
incident radiation field in the disk plane is not far from 
being isotropic, the scattered light is polarized more or 
less along the minor axis of the disk, which is the hallmark 
of the inclination-induced polarization; it is very different 
from the radial pattern seen in the face-on case (see the 
upper-left panel). In addition, both the ring of 
null polarization and the azimuthal polarization pattern in 
the outer part of the disk of the face-on case disappear, 
again because of the inclination-induced polarization. 

As emphasized in Paper I for spherical grains, the 
tendency for the inclination-induced polarization in the 
scattered light to lie along the minor axis is a simple 
consequence of the (thin) disk geometry and maximum polarization 
at $90^\circ$ scattering angle for small grains. For locations 
on the major axis of a disk of inclination angle $i$, the 
incident radiation coming from the radial direction is scattered 
by $90^\circ$ into the line of sight, whereas that from the 
locally azimuthal direction (i.e., perpendicular to the local 
radial direction in the disk plane) is scattered by  
$90^\circ + i$ or $90^\circ - i$. This difference in scattering 
angle makes the polarization from the former, which is along 
the minor axis, more important relative to that from the 
latter. Similarly, for locations along the minor axis, the 
incident light along the locally azimuthal direction is scattered 
by $90^\circ$, and that along the radial direction (in the 
disk plane) by $90^\circ + i$ or $90^\circ - i$. The difference 
increases the relative importance of the polarization from 
the former, which is again along the minor axis. This 
basic picture is qualitatively similar for both spherical and 
non-spherical grains. 

The polarization produced by direct emission is also affected by 
the disk inclination. Although the polarization vectors remain 
perpendicular to the local toroidal magnetic field projected 
onto the plane of the sky (see the middle-middle panel for the 
$i=45^\circ$ case), the polarization fraction is changed 
significantly by the inclination, especially at locations on the 
major axis, where it is reduced compared to the 
face-on case, by a factor of about 2 for $i=45^\circ$ for the 
particular grain model with $s=1.5$ adopted here 
(see Fig.~\ref{fig:p_pol}). As mentioned earlier, 
for locations on the minor axis, the aligned oblate grains 
appear ``edge-on'' to the observer independent of the inclination
angle, and their polarization fraction remains unchanged. 
Therefore, a generic feature of the polarization produced by the 
direct thermal emission of magnetically aligned oblate (or 
effectively ``oblate'', see Appendix~\ref{sec:imp&geo} for 
a discussion) grains is that, as the inclination angle $i$ 
increases, the distribution of the polarization fraction becomes 
more non-uniform azimuthally, with the radiation on the minor 
axis becoming increasingly more polarized compared to that 
on the major axis. The degree of the contrast between the 
two axes depends sensitively on the grain axis ratio $s$, 
which is unfortunately uncertain in general. 

The inclination-induced contrast between the polarizations produced 
by the direct emission on the major and minor axes is further 
increased when the scattering is also included (see the middle-right 
panel). The main reason is that, for our particular grain model, 
the polarizations produced by direct emission and scattering are 
in orthogonal directions at locations on the major axis 
(see the middle-left and middle-middle panels, see Fig.~\ref{fig:p_pol}). It 
leads to a null point at a radius $\sim 
50$~AU on the major axis where the polarization from the 
scattering cancels that from the direct emission exactly. Closer to
the center, the polarization is dominated by the scattering (which
produces a higher polarized intensity in this particular example), 
with a direction along the 
minor axis; the opposite is true beyond the null point (although this
is hard to see clearly in the middle-right panel because of low polarization
fraction). In contrast, at locations on the minor axis, the polarizations 
from both direct emission and scattering are along the same 
direction; they add to, rather than cancel, each other. The net result
is a ``butterfly-shaped'' pattern for the polarized intensity. 


%
%

Besides the strong azimuthal variation in the polarization fraction, 
there is also a significant radial dependence in the direction of
the combined polarization. At relatively small radii (within $\sim 
R_c=20$~AU), 
the polarization is dominated by scattering with
direction more or less along the minor axis. At larger radii, the
direct emission becomes more important, turning the polarization 
morphology into a more fan-like pattern. This example illustrates 
the potential richness of the interplay between the polarizations 
produced by scattering and direct emission in an inclined disk, even 
though the underlying magnetic field is simple (purely toroidal): 
the combined polarization varies both radially and azimuthally and 
in both direction and polarization fraction. In particular, it 
includes a polarization ``hole,'' where the polarizations from 
the two competing mechanisms cancel each other. We should stress 
that, for this intriguing composite pattern to appear, the 
polarized intensities from the direct emission and scattering 
must be comparable. Whether it can happen naturally is uncertain. 
In the discussion section, we will return to this and other issues, 
including the fact that the patterns of the polarization vectors 
appear very different in the scattering and emission cases for 
this intermediate inclination $i=45^\circ$, which should be
distinguished observationally. 

%
%
%

In the lower panels of Fig.~\ref{fig:ill}, we show the case of an even
more inclined disk, with $i=70^\circ$. Not surprisingly, the
inclination effect becomes more prominent for the polarizations 
produced by both scattering and direction emission. Specifically, 
the polarization from scattering has a direction nearly parallel 
to the minor axis everywhere, and a polarization fraction close 
to the maximum value of $1/3$ (see the lower-left panel). This 
pattern is similar to the highly inclined case with spherical grains,
indicating that the effect of grain non-sphericity is largely 
masked by that of inclination. For the direct emission, the
polarization near the major axis is greatly reduced relative to 
that near the minor axis (the lower-middle panel), producing a 
much more pronounced ``butterfly'' pattern than the $i=45^\circ$ 
case (the middle-middle panel). The patterns of the polarized
intensity are so distinct in the scattering and emission cases 
that one should be able to tell them apart observationally 
in principle. In practice, the characteristic ``butterfly'' 
pattern would be smeared out in disks with large inclination 
angles such as $i=70^\circ$ unless the distribution of the 
polarized intensity along the minor axis is well resolved 
spatially. Such well resolved observations should also be 
able to reveal the difference in the polarization direction 
and thus help distinguish the two cases. 
%
%

The total polarization pattern for the highly inclined $i=70^\circ$
case including both emission and scattering is shown in the 
lower-right pattern. It appears very different from that of the 
intermediate inclination ($i=45^\circ$) case (the middle-right
panel). In the $i=45^\circ$ case, the ``butterfly'' pattern in the
polarized intensity is barely recognizable for the emission only 
case, but becomes much more prominent in the combined case, 
because the polarization produced by the emission along the major 
axis is largely canceled out by that produced by the scattering. 
In contrast, in the $i=70^\circ$ case, the ``butterfly'' pattern 
is much more prominent for the emission only case, but completely
disappears in the combined case, because the low polarized 
intensity region along the major axis (the gap between the two 
``wings of the butterfly'') is filled in by the scattering-produced
polarization. In any case, the systematic change in polarization 
pattern from  $i=0^\circ$ to $45^\circ$ to $70^\circ$ is driven mainly 
by the expected decrease of the polarization from emission along 
the major axis and the increase of that from scattering at the 
same time. 
%
%

%
%
%
%

\section{The Case of NGC1333 IRAS4A1}
\label{sec:application}

%
%
%
%

IRAS4A is a well studied protobinary system in the NGC1333 region of
the Perseus molecular clouds. It is the first protostellar system
where a dust polarization pattern corresponding to an 
``hourglass-shaped'' magnetic field is detected on the 1000-AU, inner
protostellar envelope \citep{girart2006}. Given the relatively large scale (and the 
relatively low corresponding volume and column densities), it is 
unlikely for the scattering to dominate the observed polarization; the  
required grain size and column density are too large for the 
envelope. On this scale, the conventional interpretation involving 
direct emission by non-spherical grains aligned with respect to 
a (pinched) magnetic field appears secure.  

\begin{figure}
  \includegraphics[width=0.49\textwidth]{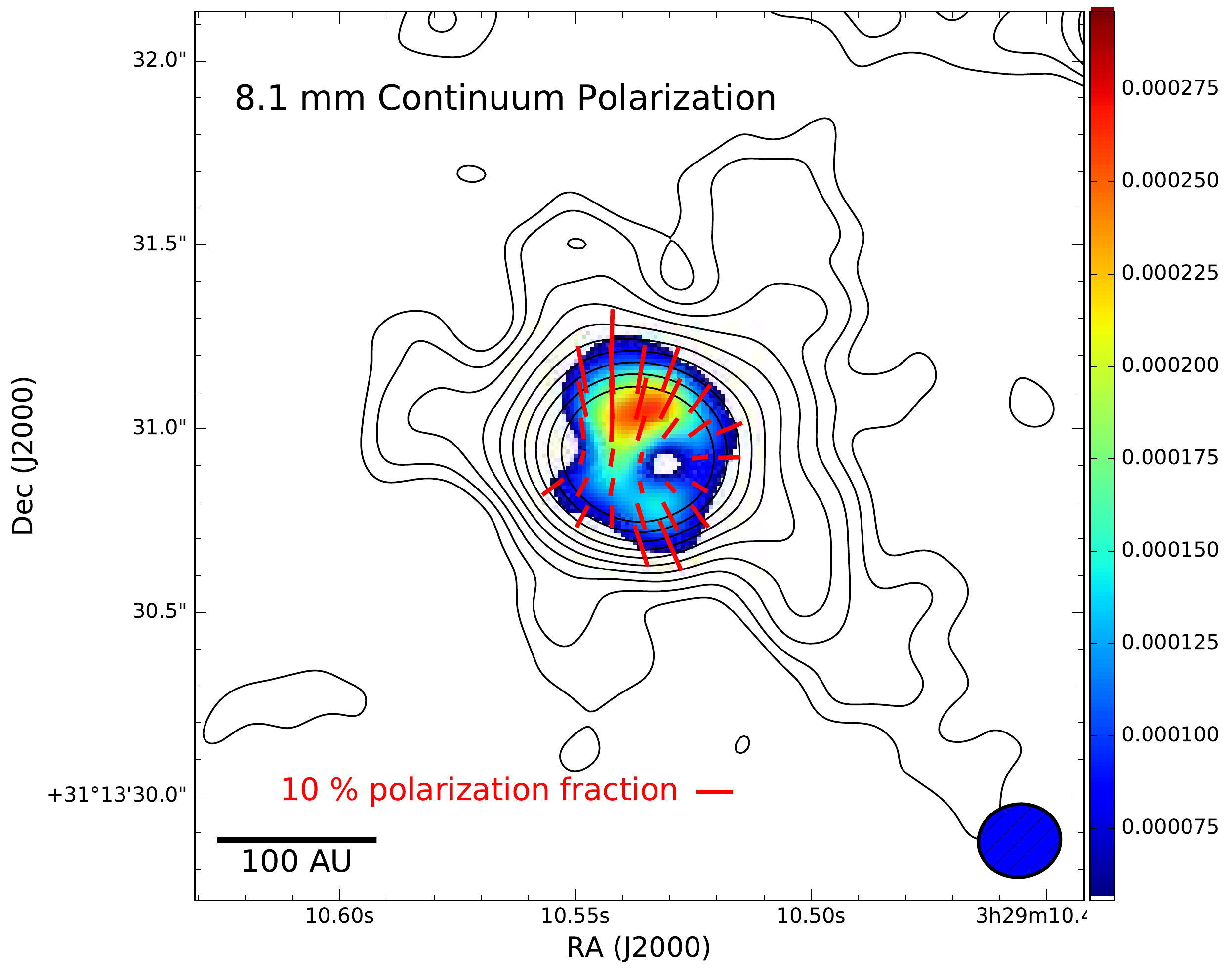}
  \caption{Polarization observed in IRAS4A1 at 8.1 mm (adapted from \citealt{cox2015}). Plotted are the total intensity (contours), polarized intensity (color map), and polarization (rather than magnetic) vectors with length proportional to the polarization fraction. The molecular outflows near the source are roughly in the north-south direction, which implies an approximately east-west orientation for the major axis \citep{santangelo2015, ching2016}.}
  \label{fig:obs}
\end{figure}

On the smaller scale of 100~AU, \cite{cox2015} recently detected
polarization at 8.1 and 10.3 mm with VLA for the brighter 
component, A1, of the protobinary system. The polarization at 
8.1 mm, which is significant for more independent beams than
that  at 10.3 mm, is reproduced in Fig.~\ref{fig:obs} for easy reference. 

As stressed by Cox et al., the polarization pattern on the 100~AU
scale appears very different from that on the 1000-AU scale. It
broadly resembles the pattern expected from direct emission by grains 
aligned with respect to a toroidal magnetic field in a face-on 
disk. It is unclear, however, whether a sizable  
rotationally supported disk exists in this source. The VLA 
continuum images appear marginally resolved, which may be 
indicative of a disk not much smaller than the resolution limit 
($\sim 50$~AU). There is, however, little kinematic data on this 
scale to confirm or reject the possibility of a Keplerian rotation. 
If the disk is indeed nearly face-on, the disk rotation would 
be difficult to measure directly. However, the red- and blue-shifted
lobes of its bipolar molecular outflows are cleanly separated 
spatially on the few 100 to few 1000 AU scale 
\citep{santangelo2015,ching2016}, indicating that the outflows are not 
exactly along the line of sight and, by implication, the disk is unlikely 
viewed face-on. 
If this interpretation is correct, the roughly north-south 
orientation of the molecular outflows would imply a disk 
major axis along approximately the east-west direction. 

\begin{figure}
  \includegraphics[width=0.49\textwidth]{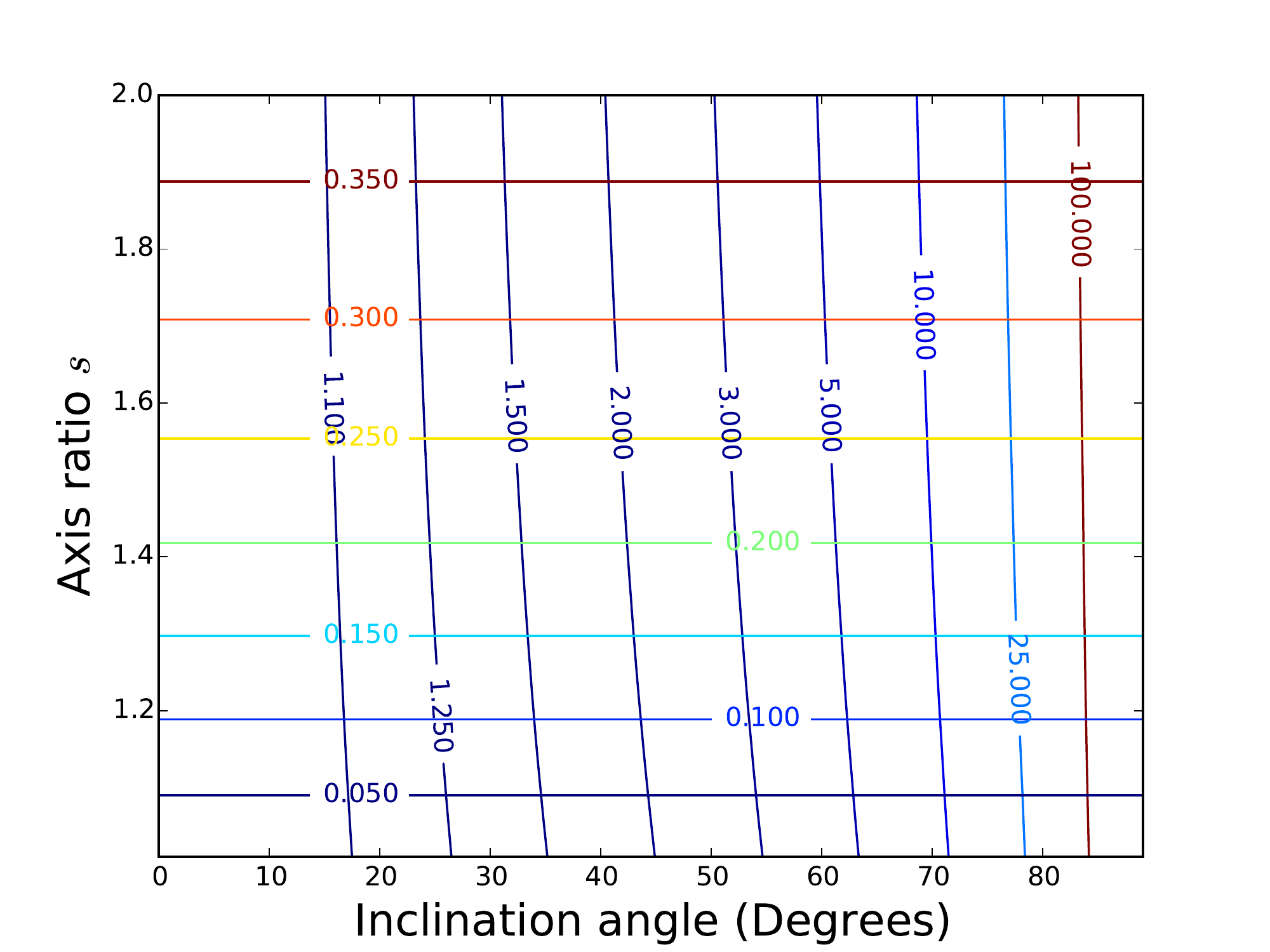}
  \caption{Lines of constant contrast $\eta$ in polarization
    fraction between the minor and major axes (the nearly
    vertical lines, with values labeled) and constant maximum
    polarization $p_{\rm max}$
    (horizontal) for direct emission in the plane of inclination angle
    $i$ and grain axis ratio $s$. Note that $\lambda$ depends weakly
    on $s$, and approaches $1/\cos^2i$ as $s\rightarrow 1$. }
  \label{fig:contrast}
\end{figure}

Additional support for an inclined disk comes from modeling of 
the 8 mm dust continuum emission, which is consistent with an 
inclination angle of $\sim 35^\circ$. Further evidence for  
significant inclination may come from the detected 
polarization pattern itself. The polarization fraction is
significantly smaller along the east-west direction than along the 
north-south direction; such a contrast is not expected in a face-on 
disk (see the upper panels of Fig.~\ref{fig:ill}). It is, however, 
qualitatively consistent with the polarization pattern produced 
by direct emission from an inclined disk with the major axis
along the east-west direction, as indicated by the molecular 
outflow orientation. As stressed earlier and illustrated 
in Fig.~\ref{fig:ill}, the polarization fraction is reduced 
along the major axis relative to that along the minor axis by
disk inclination. The magnitude of the contrast, denoted by $\eta$,
increases with the inclination angle $i$, and has a weak dependence 
on the degree of grain non-sphericity (characterized in our model by 
the grain axis ratio $s$), as illustrated in Fig.~\ref{fig:contrast}.
It is easy to show, from Equation~\ref{eq:p_abs}, that the contrast 
is given analytically 
\begin{equation}
\lambda\equiv \frac{p_{\rm abs,minor}}{p_{\rm abs,major}}\rightarrow
\frac{1}{\cos^2i}, 
\label{eq:contrast}
\end{equation}
in the limit $s\rightarrow 1$ (i.e., as the oblate spheroid approaches
a sphere, with $\alpha_3\rightarrow\alpha_1$). The above expression 
provides a good estimate for $\eta$ for the range of $s$ 
(between 1 and 2) shown in Fig.~\ref{fig:contrast}. 

Also plotted in the figure are lines of constant maximum 
polarization fraction $p_{\rm max}$. This maximum value depends 
on the grain axis ratio $s$ but not the inclination angle, and 
is reached at locations along the minor axis (i.e., $p_{\rm
  abs,minor}=p_{\rm max}$). This diagram can help evaluate 
whether the polarization observed in a particular source comes 
from direct emission or not. 

\begin{figure}
  \includegraphics[width=0.49\textwidth]{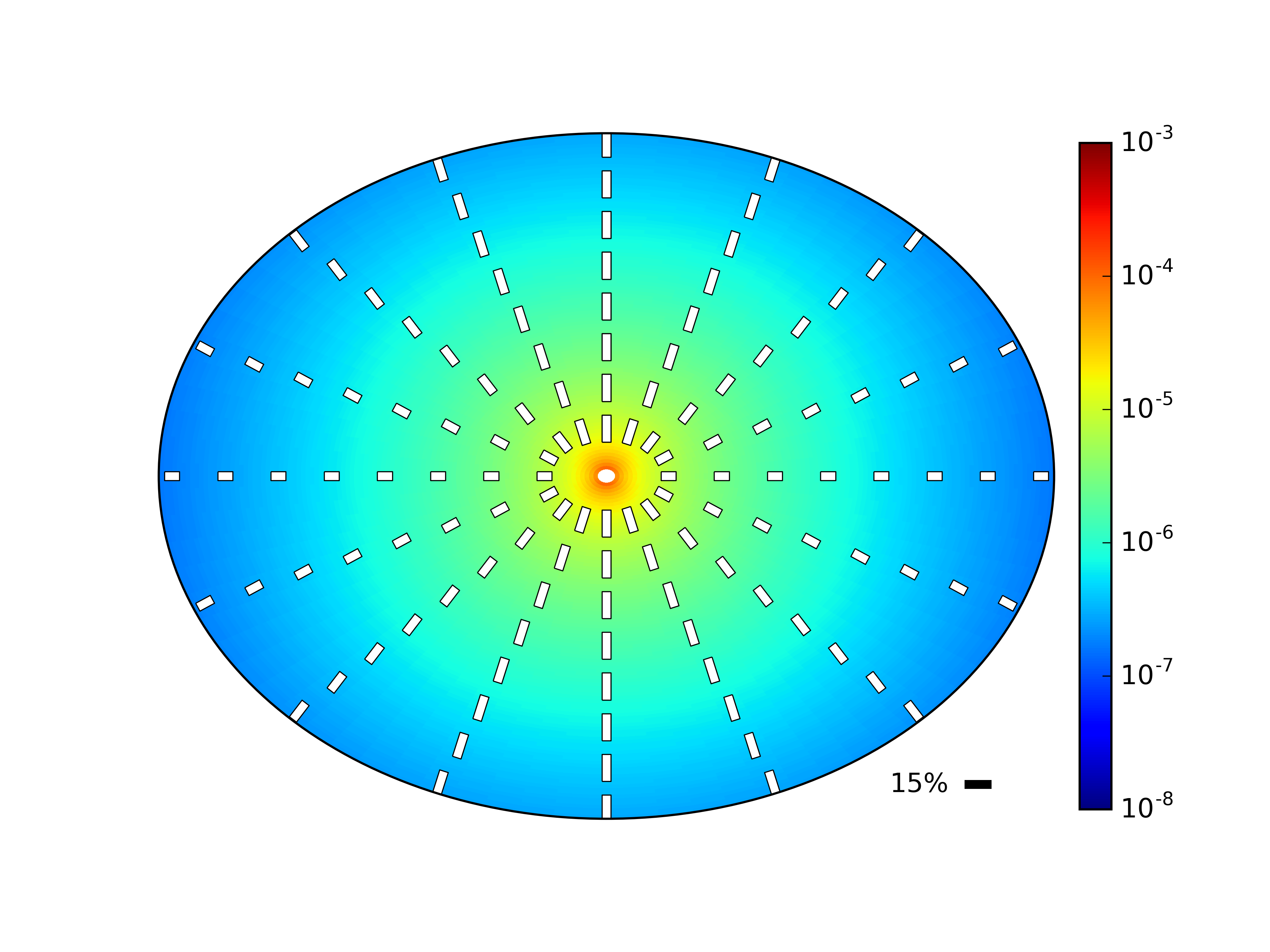}
  \includegraphics[width=0.49\textwidth]{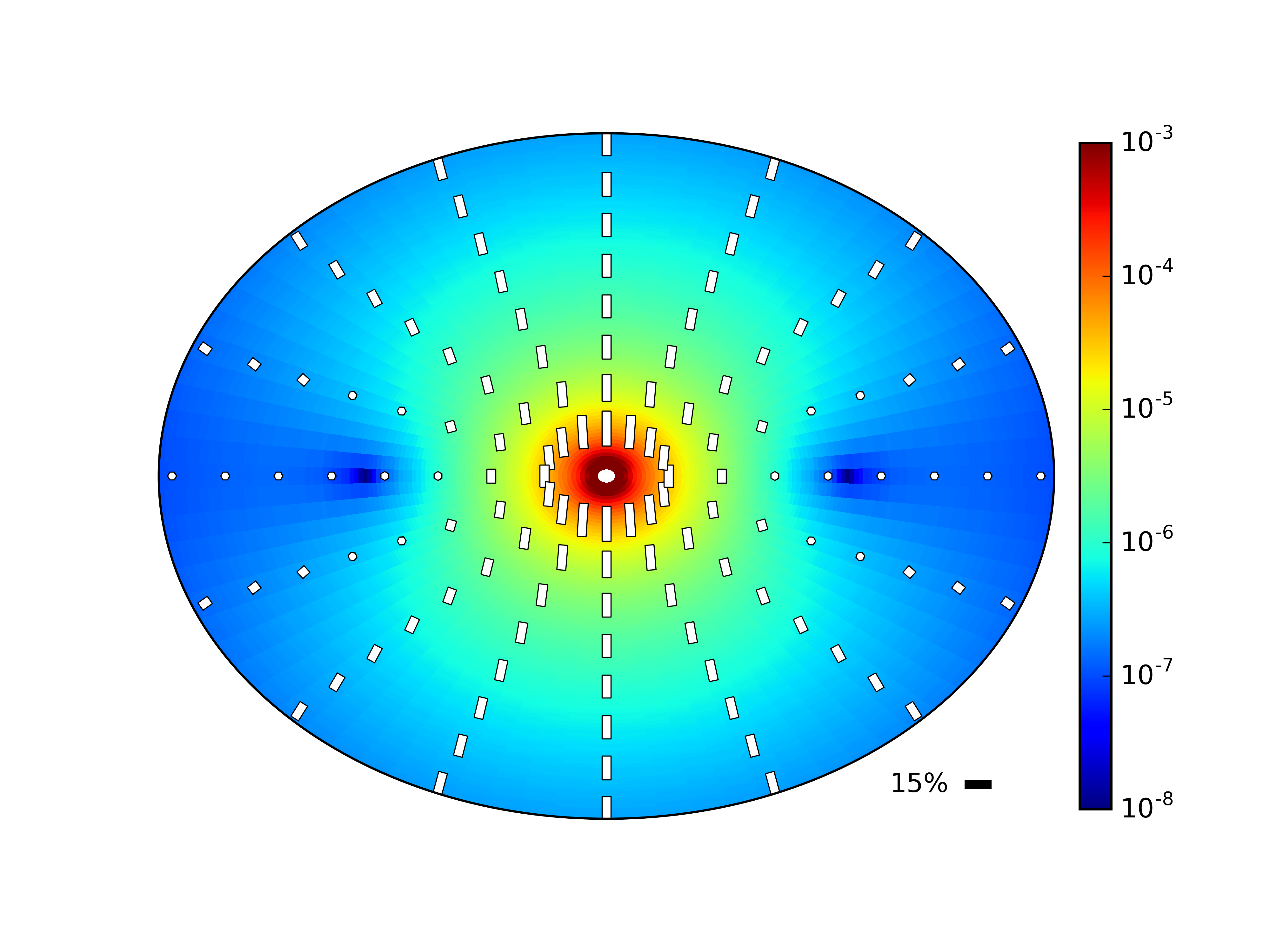}
   \caption{Polarization models with and without scattering. Plotted are the polarized intensity (in units of $B_\nu(T_0)$, color map) and polarization vectors with length proportional to polarization
    fraction for emission only (upper panel) and for both emission and scattering (lower panel). The lower panel resembles the observed polarization in IRAS4A1 shown in Fig.~\ref{fig:obs} more closely than the upper panel (see text for discussion). 
}
  \label{fig:app}
\end{figure}

In the case of IRAS4A1, the polarization fraction is the highest 
along the minor axis in the north-south direction 
(see Fig.~\ref{fig:obs}), consistent with the direct emission
interpretation. The maximum value in the north is $\sim 18\%$, which
is somewhat larger than that in the south ($\sim 12\%$). In the 
grain model adopted in this paper, these degrees of polarization 
correspond to a grain axis ratio $s$ of $\sim 1.4$ and  
$\sim 1.2$ respectively in this interpretation. 
The upper limit on the inclination angle was set by fitting 
the 8 mm continuum data to a disk model in the \textit{uv}-plane, 
following the method used in \cite{segura-cox2016}.  The 
shortest baselines ($<$ 350 k$\lambda$) were omitted from 
the data to better exclude envelope emission for the modeling.  
Since the inclination angle is likely less than $\sim 45$ degrees 
based on the continuum modeling\footnote{\protect\cite{ching2016}
suggested a larger inclination
  angle of $\sim 70-80^\circ$ based on outflow modeling, although the 
  inferred angle depends strongly on their model assumptions. 
If the inclination is indeed
  this high, the scattering would be more important relative to direct
  emission.}, the contrast $\eta$ should be less than $\sim 1/\cos^2
45^\circ=2$. This expectation is confirmed in the left panel of 
Fig.~\ref{fig:app}, where we show the polarization pattern at 8~mm 
from emission by perfectly aligned oblate grains of 0.6~mm in size 
and $s=1.3$ in axis ratio (adopting the same grain material 
as in \S~\ref{sec:disk}, which has a complex dielectric constant 
$\epsilon = 3.78+0.0075j$ at 8~mm). The inclination angle was 
set to $i=40^\circ$, which is on the high side of the range preferred 
by the continuum modeling. As expected, there is some contrast 
between the minor and major axes in the polarization fraction (and
polarized intensity). The contrast appears less than that suggested 
by observation: roughly 12-18\% along the (minor) north-south axis 
and approximately 3-4\% along east-west. That is, the contrast
$\eta$ is at least a factor of 3, and likely significantly 
higher. In order to produce such a high contrast, a disk 
inclination angle of $\sim \arccos(1/\sqrt{3})\approx 55^\circ$ or more 
is needed according to Equation~(\ref{eq:contrast}). Such a
large inclination, although cannot be ruled out completely, is
unlikely based on the continuum modeling. 

Another, perhaps more severe, drawback of the emission only model is 
that it predicts a purely east-west orientation for the polarization 
vectors on the major axis, which
matches the observed vectors near the western edge but not those 
closer  to the center, which are oriented more or less north-south (i.e., 
along the minor axis). 
The orientations of these central vectors can naturally arise from 
scattering, which has the added advantage of canceling out some of
the polarization produced by emission on the major axis and 
thus bringing the contrast $\eta$ closer to the observed level. 
This is illustrated in the right panel of Fig.~\ref{fig:app}, 
where we include the contributions to the polarization from both
emission and scattering. In this particular example, the scattering
dominates the emission near the center and visa versa near the 
edge. Two polarization ``holes'' are produced at a distance of 
$\sim 25$~AU along the 
major axis, one on each side of the origin. They broadly resemble 
the polarization ``hole'' to the west of the center\footnote{We
  checked that the polarization ``hole'' is not where the 
emission at longer wavelengths (1 and 4 cm) peaks, and is therefore
unlikely caused by unpolarized free-free emission.} and, to a 
lesser extent, the low-polarization ``bay'' to the east. The 
inclusion of scattering appears to have improved the agreement 
between the model and observations significantly, at least in some 
broad features. 

The inclusion of scattering does not improve the agreement in other 
observed features, however.  For example, the north-south 
asymmetry in the polarized intensity (see Fig.~\ref{fig:obs}) 
cannot be accounted for in our simple semi-analytic model that 
assumes an axisymmetric disk structure. Asymmetry in the 
disk properties, such as the dust distribution, could be a culprit.  
 Another discrepancy is that the polarized 
intensity is peaked at the center in the model but not in the 
observed map. However, the central region may be  
optically thick, which would reduce the polarization fraction for
both the directly emitted and scattered light (Liu et al. 2016). 
In any case, more
detailed models will be needed to explain these features, 
especially when they become better quantified with higher 
resolution and sensitivity observations in the future. 

\section{Implications and Future Refinements}
\label{sec:discussion}

We have found suggestive evidence that the dust scattering may have 
contributed significantly to the polarization observed in NGC1333 
IRAS4A1 on the 50~AU scale, especially in the central region and 
along the major axis. However, the concentration of the polarized 
light along the minor axis and the ``fanning out'' of most of the
polarization vectors point to a polarization pattern dominated by 
the direct emission from grains aligned with respect to a toroidal 
magnetic field as the dominant mechanism, especially in the outer 
regions, with the strong implication that the disk is indeed 
magnetized. This is very different from the case of HL Tau disk, 
where the polarized light is concentrated along the major axis, 
and all polarization vectors are more or less parallel to the minor 
axis \citep{stephens2014}. As emphasized in Paper I (see also 
\citealt{kataoka2015b}), these features are explained more 
naturally by dust scattering than direct emission. These two 
examples illustrate the diversity of the polarization 
pattern on the disk scale and the need to include both scattering 
and direct emission for interpreting the observations. The need 
will only increase in the near future as ALMA disk polarization 
observations with higher spatial resolution and sensitivity become 
available. 

\begin{figure}
  \includegraphics[width=0.49\textwidth]{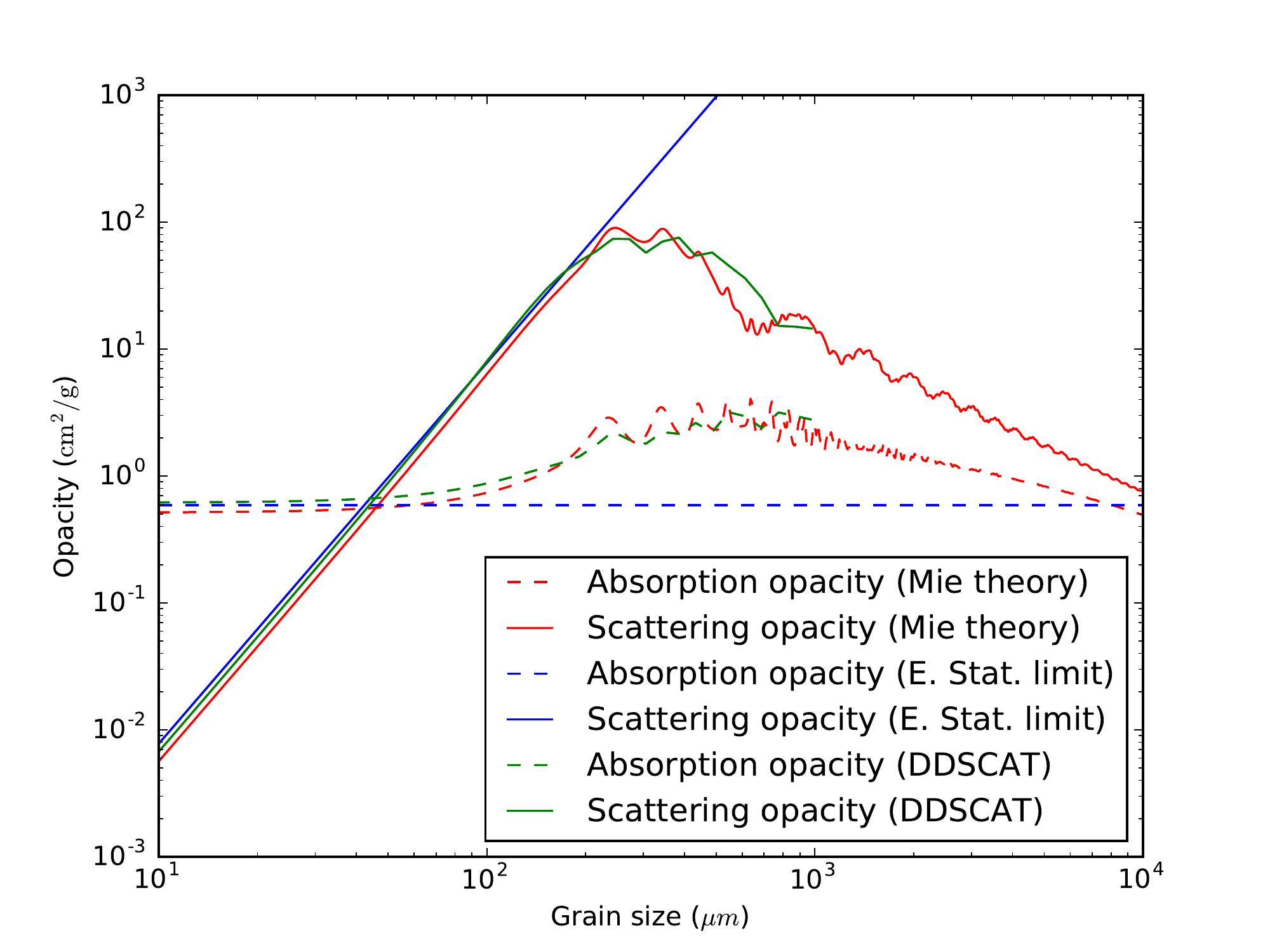}
   \caption{Scattering (solid line) and absorption (dashed) opacities at 1 mm as a function of grain
     size for oblate grains with $s=1.5$ computed using the discrete
     dipole approximation (green lines) and under the electrostatic approximation for small particles (blue lines).  Note that the scattering opacity obtained under the electrostatic approximation is valid up to a grain size of $\sim 0.2$ times the wavelength $\lambda$, 
and it exceeds the absorption opacity for grains larger than $\sim
0.05 \lambda$. Opacities computed from Mie theory for spherical grains
     of the same size are also shown (red lines) for comparison.} 
  \label{fig:opacity}
\end{figure}

There are several factors that determine the relative importance of
the scattering and emission in disk polarization, including the 
grain properties, disk structure and inclination. A key factor is 
the grain size, to which the scattering opacity $\kappa_{sca}$ is highly
sensitive. This sensitivity is illustrated in Fig.~\ref{fig:opacity},
where we plot the scattering and absorption opacities as a function 
of the grain size $r_e$ for oblate grains with
an axis ratio $s=1.5$, obtained using both the electrostatic 
approximation and discrete dipole approximation \citep[DDSCAT]{DF94} 
at wavelength $\lambda=1$~mm. 
Also plotted for comparison is the opacity for spherical grains of the
same size computed from the Mie theory. 
As mentioned earlier, the scattering opacity $\kappa_{sca}
\propto {r_e}^{3}$ for grains smaller than about $\lambda/(2\pi)$. It
starts to exceed the absorption opacity $\kappa_{abs}$ only for grains
larger than $\sim 0.05 \lambda$. The sensitive dependence of
$\kappa_{sca}$ on $r_e$ is a double-edged sword. It implies a
relatively narrow range in grain size, from $\sim 0.05 \lambda$ to 
$\sim 0.2 \lambda$, for the scattering to be competitive with direct 
emission and the electrostatic approximation adopted in this paper 
to hold\footnote{Note that the grain sizes used in \S~\ref{sec:disk} 
and
  \S~\ref{sec:application} are in this range, so our treatment is
  self-consistent.}. Scattering may still dominate direct emission 
for grains above this size range, but its polarization patterns will
likely be quite different from those discussed in this paper
(including, e.g., polarization reversal, Paper I) and will 
need more elaborated methods, such as the Discrete Dipole Approximation 
\citep[e.g.,][]{DF94}, to determine; we will postpone such a 
treatment to a future investigation.

On the other hand, if the polarization pattern observed in a disk
requires dust scattering to explain, the size of the scattering 
grains must lie in a relatively narrow range. The case of IRAS4A1 
is particularly interesting in this context. To produce significant 
polarization at 8~mm by dust scattering, the grains must be 
roughly millimeter-sized (or larger). In this source, there is 
evidence for polarization from direct emission as well. If 
the polarized emission is dominated by the same grains that are 
responsible for the scattering, it would imply that large, 
millimeter-sized, grains can indeed be aligned with respect to 
the magnetic field inside the disk. This inference is important 
because, compared to the micron-sized (or smaller) grains that 
are more commonly discussed in the grain-alignment literature, the 
much larger, millimeter-sized, grains are more difficult to align 
by radiative torque because of their slower internal relaxation 
\citep{HL09} and slower Lamor precession around the 
field \citep{lazarian2007}. The latter obstacle can in principle be 
overcome with a strong enough magnetic field. Therefore, alignment 
of large grains can potentially provide an indirect estimate of 
the lower limit to the field strength that is all-important to 
the disk dynamics; we will postpone the quantification of this 
limit to a future investigation. 

A potential complication is that the grains responsible for the
scattering and direct emission may not have the same sizes. For 
example, in the case of IRAS4A1, the central part of the disk where 
scattering appears to dominate the polarization may have large 
grains while the direct emission-dominated outer part could have 
smaller grains. Indeed, there is evidence for such a spatial 
gradient, with the grain size increasing toward the center, from 
the distribution of opacity spectral index $\beta$ in a number 
of (relatively evolved) disks \citep[e.g.,][]{perez2012,testi2014,guidi2016}. 
The gradient is also expected 
on theoretical grounds \citep[e.g.,][]{birnstiel2012}. The inward 
increase in grain size tends to make the scattering-induced 
polarization more important at smaller radii (in addition to 
a higher column density there), as appears to be the case in 
IRAS 4A, although the optical depth close to the 
center could be substantial, which may invalidate the optical 
thin approximation and single scattering assumption adopted in 
the paper. These effects should be treated self-consistently 
in more refined models in the future, together with the expected 
spatial variation of grain properties. Another refinement is to
  include the polarization of the incident light in treating the
  scattering. 

%
%
If the observed polarization is dominated by direct emission from 
magnetically aligned grains, the polarization fraction may provide 
a handle on the grain shape. For perfectly aligned oblate
spheroids, there is a one-to-one relation between the grain axis 
ratio $s$ and the maximum polarization fraction $p_{max}$ (see
Fig.~\ref{fig:contrast}). For example, values of $p_{max}=15\%$ 
and $30\%$ would imply axis ratios of $s\approx 1.3$ and $1.7$, 
respectively. However, the polarization could also be produced by
prolate grains, whose optical properties are similar to those 
of the oblate grains when averaged around the magnetic field 
direction (see Appendix~\ref{sec:imp&geo}). Furthermore, alignment  
with the magnetic field may not be perfect, especially for large
grains with Larmor precession time scales longer than the disk 
lifetime. For imperfectly aligned grains, larger deviation from
spherical shape is needed to produce the same degree of polarization. 
Therefore, there is a degeneracy between different grain shapes 
(oblate vs prolate) and between the grain shapes and their degrees 
of alignment that is difficult to break with the observed 
polarization fraction alone. Grain growth models and detailed 
grain alignment calculations, together with higher resolution 
and sensitivity data, may be needed to break the degeneracy. 

\section{Conclusion}
\label{sec:conclusion}
Using the electrostatic approximation, we have taken a first step 
toward developing a general theory for disk polarization in 
millimeter and centimeter that includes both direct emission from 
magnetically aligned, non-spherical grains and scattering by the 
same grains, with an emphasis on the relative importance of these 
two mechanisms and how they are affected by disk inclination. 
We have adopted the approximation of unpolarized incident
light for scattering, which could affect the polarization produced
by scattering at a level up to a few tens of percent (see
Appendix \ref{sec:unpolarized}). With this caveat in mind, the main
results are as follows:

1. The polarizations produced by scattering and direct emission both
depend strongly on the disk inclination, which changes the relative
importance of the two, especially along the (projected) disk major
axis in the plane of the sky. This change was illustrated analytically
with a simple case where oblate grains are perfectly aligned with a
purely toroidal magnetic field at a location on the major axis where
the incident radiation field is assumed  isotropic (see Fig.~\ref{fig:p_pol}). For a
nearly face-on disk, both scattering and direct emission produce
polarization along the major axis (or radial direction) at the
location; they tend to reinforce each other. As the inclination angle
$i$ increases, the direction of the scattering-induced polarization
switches to the minor axis, with the polarization fraction increasing
to $1/3$ as $i\rightarrow 90^\circ$. In contrast, the polarization
produced by direct emission remains along the major axis, with the
polarization fraction decreasing monotonically to zero as
$i\rightarrow 90^\circ$. Therefore,  for large disk inclinations, the
polarizations from scattering and direct emission tend to cancel each
other on the major axis, with the scattering dominating the direct
emission in the limit of edge-on disks. For less extreme disk
inclinations, the relative importance of the two competing mechanisms
depends on the properties of the dust grains, especially their size
and degree of non-sphericity, and the ratio of the Planck function
$B_\nu(T)$ for thermal dust emission and the mean intensity $J_\nu$ of
the incident radiation field to be scattered by the grains. 

%
%
2. The scattering and direct emission by magnetically aligned,
non-spherical grains produce polarization patterns that should be easily
distinguishable in general but not always. This was illustrated with a
geometrically and optically thin dust disk of a prescribed column
density and temperature distribution and a purely toroidal magnetic
field (see Fig.~\ref{fig:ill}). For significantly inclined disks, the 
difference
between the two mechanisms is most pronounced at locations on 
the major axis, where the polarized intensity is enhanced relative to that
on the minor axis and the polarization direction is along the minor
axis for scattering while the opposite is true for direct emission. 
For nearly face-on
disks, the direction of the scattering-induced polarization near the
disk center where the radiation field is more or less isotropic in the
disk plane is the same as that from direct emission, making it hard to
distinguish the two (both radial). At larger radii where the radiation
field in the disk plane is more radially beamed, the
scattering-induced polarization switches to the azimuthal direction,
which is orthogonal to that from the emission. The interplay between 
these two competing mechanisms can yield interesting new polarization
patterns, especially when their polarized intensities are
comparable. Particularly intriguing is the pattern produced in a disk
of intermediate inclination with the scattering dominating the inner
region of the disk and the emission the outer:  the polarization
directions are nearly uniform (along the minor axis) at small radii,
and become increasingly radial at larger distances, with two ``null''
points located on the major axis (one on each side of the origin)
where the polarizations from scattering and direct emission cancel out
exactly. The ``null'' points serve as a signpost for both mechanisms
contributing significantly to the polarization.   
%
%

3. There is suggestive evidence that the polarization pattern observed
in NGC1333 IRAS4A1 at 8~mm is shaped by a combination of direct
emission and scattering. The scattering and direct emission naturally 
account for, respectively, the relatively uniform polarization 
directions observed in the central region and the roughly radial 
pattern at larger distances (see Fig.~\ref{fig:obs}). Most
interestingly, there is clear evidence for at least one ``null'' 
point in the observed polarization map, which can naturally be
interpreted as the location on the major axis of an inclined 
disk where the polarizations from the scattering and direct 
emission cancel each other. The implied disk orientation matches 
that required for launching the observed molecular outflows. 

%
%
4. If both direct emission and scattering from the same magnetically aligned 
grains indeed contribute significantly to the polarization observed 
in IRAS 4A1, it would imply not only that a magnetic field exists 
on the disk scale, but that it is strong enough to align large, 
possibly millimeter-sized, grains, at least in this source, with 
potentially far reaching consequences for the disk dynamics and 
evolution. This inference remains tentative,  
however, in this early stage of observations and modeling of disk 
polarization. The situation should be greatly improved in the near 
future with the higher resolution and sensitivity ALMA observations 
and model refinements. 

\vskip 1cm 

We thank Chat Hull and an anonymous referee for helpful comments.
ZYL and HFY are supported in part by NASA NNX14AB38G and NSF AST-1313083.




\appendix
\section{Prolate grains and imperfect alignment}
\label{sec:imp&geo}

In \S~\ref{subsec:oblate1} and \ref{subsec:oblate2}, 
we have considered in detail only oblate grains. For
non-oblate grains that have their shortest axes aligned with the 
local magnetic field, the situation is qualitatively similar to the 
oblate grain case, as a result of either rapidly grain rotation around
the field line or averaging over an ensemble of grains. For example, 
consider prolate grains with the semi-diameters $a_1>a_2 = a_3$ and
intrinsic polarizability $|\alpha_1| > | \alpha_2| = |\alpha_3|$. 
Let the minor axis $a_3$ be aligned with the magnetic field. The
polarizability along the field direction remains unchanged (i.e., 
$\alpha_{\parallel,3}=\alpha_3$, where the subscript $\parallel$ 
denotes "parallel" to the local magnetic field ), whereas the two components
perpendicular to the field become $\alpha_{\perp,1}=\alpha_{\perp,2} 
= (1/2)(\alpha_1+\alpha_2)$, which is the average over the azimuthal 
angle around the field line \citep[see e.g.,][]{LD85}.
Therefore, the effective (averaged) 
polarizabilities for the prolate grains become
$|\alpha_{\perp,1}|=|\alpha_{\perp,2}| > |\alpha_{\parallel, 3}|$, 
which have the same ordering as
the oblate grain case. In other words, the averaging makes the 
prolate grains behave effectively as ``oblate'' grains as far 
as the polarization is concerned, although their efficiency in 
producing polarization is reduced somewhat compared to the 
oblate grains that have the same long-to-short axis ratio 
\cite[see, e.g., ][]{HD95}.

Another potential complication is that the grains may not be perfectly
aligned with respect to the magnetic field. For example, it is likely
for the grains to wobble around the field line 
\citep[see e.g.][]{HL12}. The wobbling is expected to be more 
important for larger grains, since their alignment is made less 
efficient by the longer Larmor precession time. Determining the 
degree of alignment requires a detailed study of the grain alignment 
mechanism, which is beyond the scope of this paper. Here we 
illustrate the effects of imperfect alignment through parametrization.

For simplicity, let us consider oblate grains with the symmetric axis
along the shortest axis $a_3$. Let the grain's shortest axis wobble
around the local magnetic field, which is fixed in space, with an
instantaneous polar angle $\theta$ and azimuthal angle $\phi$ with
respect to the field direction. With a simple frame rotation,
we can obtain the polarizability matrix in the lab frame, i.e., the frame 
fixed with respect to the magnetic field (rather than the wobbling
grains). Since the system is symmetric with respect to the field direction, 
we can average over the azimuthal angle $\phi$, which leaves the 
elements of the polarizability matrix in the lab frame depending 
only on the polar angle $\theta$: 
\begin{equation}
  \begin{split}
  \bar{\alpha}= \mathrm{diag}\Bigg\{
\frac{1}{2}(\alpha_1+\alpha_3)+\frac{1}{2}(\alpha_1-\alpha_3)\mean{\cos^2(\theta)}, \\
\frac{1}{2}(\alpha_1+\alpha_3)+\frac{1}{2}(\alpha_1-\alpha_3)\mean{\cos^2(\theta)}, \\
\alpha_1 - (\alpha_1-\alpha_3)\mean{\cos^2(\theta)}\Bigg\}
  \end{split}
  \label{eq:alpha_imp}
\end{equation}
where $\mathrm{diag}\{\}$ represents a diagonal matrix and 
$\mean{\cos^2(\theta)}$ is an ensemble average. We can see
that the matrix preserves the form of polarizability matrix of 
oblate grains with two equal components bigger than the third one. 
When $\mean{\cos^2(\theta)} = 1$, we recover the perfect 
alignment result. In the opposite limit of completely random 
grain orientation, we have $\mean{\cos^2(\theta)}=1/3$, which yields  
$\bar{\alpha}=(1/3)(2\alpha_1+\alpha_3)\bar{I}$, where $\bar{I}$ 
is the identity matrix. As expected, there would be no polarization
from direct dust emission in this case, and the polarization would 
be completely dominated by scattering. This limiting case is an
example of the general trend that imperfect grain alignment tends 
to increase the importance of scattering relative to direct 
emission. 

\begin{figure}
  \includegraphics[width=0.49\textwidth]{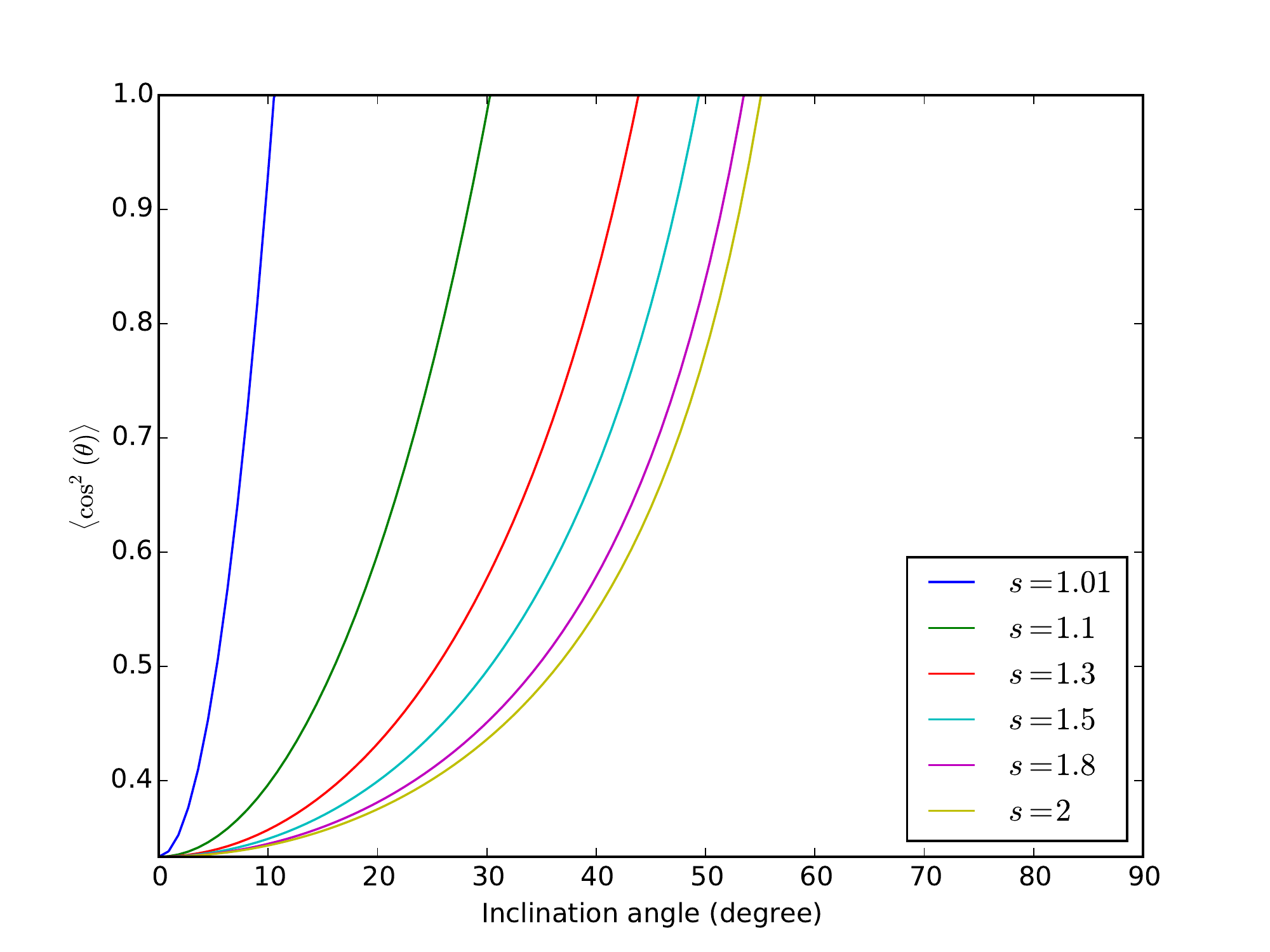}
  \caption{Effects of imperfect grain alignment, parametrized by the
    value of $\mean{\cos^2(\theta)}$, on the relative importance of
    scattering and direct emission for polarization for the case of  
    $\sigma_{\rm abs}B_\nu/\sigma_{\rm sca}J_\nu=2$. For each value of
    axis
    ratio $s$, the polarization is dominated by direct emission in the
    parameter space to the upper-left of the corresponding curve, and 
    by scattering to the lower right. }
  \label{fig:ori}
\end{figure}

To illustrate the above trend further, we consider how imperfect grain 
alignment, as parametrized by the value of $\mean{\cos^2(\theta)}$, 
affects the transition inclination angle $i_t$ (discussed in \S~\ref{subsec:oblate2} 
and Fig.~\ref{fig:rratio}) where the polarization 
produced by scattering cancels that from direct emission completely,
for the fiducial value of the ratio $\sigma_{\rm abs}B_\nu/\sigma_{\rm
  sca}J_\nu$. The results are shown in Fig.~\ref{fig:ori}. Clearly, for each value
of the axis ratio $s$, the scattering starts to become important at 
a smaller inclination angle as the grain alignment becomes worse
(i.e., as the parameter $\mean{\cos^2(\theta)}$ decreases). Another
way to interpret the curve for each $s$ in the figure is that, in 
order for the direct emission to dominate the total polarization, two 
conditions must be satisfied: (1) the inclination angle $i$ must 
be less than a critical value (the value of the transition angle 
$i_t$ in the perfectly aligned limit), and (2) the grains must be 
sufficiently aligned so that the parameter $\mean{\cos^2(\theta)}$ 
is larger than the value at the intersection of the curve and the  
vertical line passing through the angle $i$). 

In summary, in the presence of a magnetic field, the local field direction 
serves as a symmetry axis for the system. Averaging around this axis
makes non-oblate grains behave effectively as oblate grains regardless
of their shape and degree of alignment. It provides a strong
motivation to concentrate on oblate grains with different values of
axis ratio $s$, since the results in the more general cases will be
qualitatively similar. The downside of the averaging is that there is
a strong degeneracy between the degree of alignment, characterized by 
the quantity $\mean{\cos^2(\theta)}$, and the degree of the grain 
non-sphericity, characterized by $s$. In particular, imperfectly 
aligned ``needles'' might have similar optical properties as 
perfectly aligned ``pancakes,'' making it difficult to tell them apart 
based on polarization observations. 

\section{Approximation of unpolarized incident light for scattering}
\label{sec:unpolarized}

Here we evaluate the effect of the approximation of
unpolarized incident light on the polarization produced by scattering.
For a disk with a purely 
toroidal magnetic field, the incident radiation will be 
polarized along the $z$ direction, i.e., the normal direction of the 
disk (see the Cartesian coordinate system defined in the second
paragraph of \S~\ref{subsec:oblate1}), so that is Stokes parameters 
$U=V=0$. In this case, the polarization fraction of the scattered 
light can be estimated roughly as:
\begin{equation}
  p \sim \frac{\mean{I}\mean{S_{21}} + \mean{Q}\mean{S_{22}}}
  {\mean{I}\mean{S_{11}}+\mean{Q}\mean{S_{12}}} 
  \sim \frac{\mean{S_{21}}}{\mean{S_{11}}} 
  \frac{1+\tilde{p} \frac{\mean{S_{22}}}{\mean{S_{12}}}}
  {1+ \tilde{p} \frac{ \mean{S_{12}} }{ \mean{S_{11}} } }
\label{eq:correction}
\end{equation}
where $\tilde{p}\equiv \mean{Q}/\mean{I}$, and the brackets 
denote angle-averaging. The dust polarization fraction observed 
in young star disks is of order $\sim 10\%$ \citep{cox2015} or 
less (typically of order $1\%$; \citealt{stephens2014}). 
If such low values are representative of the polarization 
fraction of the direct thermal emission, we would expect 
$\tilde{p}$ to be of this order as well, i.e., $\tilde{p}\sim 
1-10\%$. The factor $\mean{S_{12}}/\mean{S_{11}}$ in the 
denominator of the above equation is of the same order 
as $\tilde{p}$, so we expect the correction term 
$\tilde{p} \frac{ \mean{S_{12}} }{ \mean{S_{11}} }$ in the 
denominator to be of order $\tilde{p}^2\sim 10^{-2}-10^{-4}$, 
which is negligible.

The correction term in the numerator of equation~(\ref{eq:correction})
is expected to be larger, because the ratio $\mean{S_{22}}
/\mean{S_{12}}$ is typically of order a few (rather than the 
much smaller $\tilde{p}$). It is expected to affect the intensity 
of the scattering-produced polarized radiation at a few to a few 
tens of percent level. 

We do not expect the approximation of unpolarized
incident light to significantly affect the polarization pattern 
produced by scattering, especially in the central region of an
axisymmetric disk, where the incident radiation is nearly 
isotropic in the disk plane. In this case, the same 
angle-averaging as in Section \S~\ref{subsec:oblate1} 
yields $\mean{S_{32}}=\mean{S_{42}}=0$, which implies that the 
scattering of incident light polarized perpendicular to the 
disk will not produce any $U$ or $V$ component, just as in 
the case of unpolarized incident light.

\bsp	
\label{lastpage}
\end{document}